\documentclass[iopart,twocolumn,groupeaddress, superscriptaddress,showpacs]{revtex4}
\usepackage{bbm}
\usepackage{epsfig}
\usepackage{array}
\usepackage{braket}
\usepackage{float}
\usepackage{lmodern,amssymb}
\usepackage[dvipsnames]{xcolor}
\usepackage{lipsum}
\usepackage{amsmath}
\usepackage{latexsym}
\usepackage{amssymb}
\usepackage{lipsum}
\usepackage[english]{babel}
\usepackage{mathtools}
\usepackage{braket}
\usepackage{graphicx}
\usepackage[caption=false]{subfig}
\usepackage{wrapfig}
\usepackage{dsfont}
\usepackage{etoolbox}
\usepackage{comment}
\usepackage[colorlinks=true, citecolor=blue, urlcolor=blue]{hyperref}
\usepackage{float}
\usepackage{amsfonts}
\usepackage{soul}
\usepackage[T1]{fontenc}

\newcommand{\beq}[0]{\begin{equation}}
\newcommand{\eeq}[0]{\end{equation}}
\newcommand{\non}{\nonumber}
\def\be{\begin{equation}}
\def\ee{\end{equation}}
\def\bea{\begin{eqnarray}}
\def\eea{\end{eqnarray}}
\newcommand{\ba}{\begin{eqnarray}}
\newcommand{\ea}{\end{eqnarray}}
\usepackage{hyperref}

\usepackage[bb=boondox]{mathalfa}
\usepackage[utf8]{inputenc}
\usepackage{textcomp}

\begin{document}
\title{Fluctuations and optimal control in a Floquet Quantum Thermal Transistor}

\author{Samir Das }\email{dassamir579@gmail.com}
\affiliation{Department of Chemistry, Indian Institute of Technology Kanpur, Kanpur 208016, Uttar Pradesh, India}
\affiliation{Department of Physics, Universidad de Santiago de Chile, Av. Victor Jara 3493, Santiago, Chile}
\affiliation{Millennium Institute for Research in Optics, Concepci´on, Chile}
\author{Shishira Mahunta,}
\email{shishiram@iiserbpr.ac.in}
\affiliation{Department of Physical Sciences, IISER Berhampur, Berhampur 760010, Odisha, India}
\author{Nikhil Gupt}
\affiliation{Department of Chemistry, University of Pennsylvania, Philadelphia, Pennsylvania 19104, USA}
\author{Victor Mukherjee}\email{mukherjeev@iiserbpr.ac.in}
\affiliation{Department of Physical Sciences, IISER Berhampur, Berhampur 760010, Odisha, India}
\author{Arnab Ghosh}\email{arnab@iitk.ac.in}
\affiliation{Department of Chemistry, Indian Institute of Technology Kanpur, Kanpur 208016, Uttar Pradesh, India}

\begin{abstract} 
We use Full Counting Statistics to study fluctuations and optimal control in a three-terminal Floquet quantum thermal transistor. We model the setup using three qubits (termed as the emitter, collector, and base) coupled to three thermal baths. As shown in Phys. Rev. E ${\bf 106}$, 024110 (2022), one can achieve a significant change in the emitter and collector currents through a small change in the base current, thereby achieving thermal transistor operation. We quantify fluctuations using variance, Fano factor and Thermodynamic Uncertainty Relation (TUR). We have proposed analytical  lower bounds to the Fano factor and TUR ratios, and verified the same using  examples of sinusoidal and $\pi$-flip modulations. We  also use optimal control through the Chopped Random Basis optimization protocol in order to significantly enhance the amplification obtained in the transistor, or alternatively, to reduce the Fano factor.  Importantly, we identify regime of ideal operation in the parameter space, characterized by high amplification and low Fano factor, as well as regime of worst performance, associated with large Fano factor, without a substantial enhancement in the amplification. Interestingly, optimization protocol allows us to reach the analytical lower bound of the Fano factor. We expect our results will be highly relevant for developing heat modulation devices in near-term quantum technologies. 
\end{abstract}


\maketitle
\section{INTRODUCTION}
The study of quantum thermodynamics focuses on thermodynamic phenomena in the quantum realm, and is crucial for the development of quantum thermal devices, such as quantum heat engines \cite{cangemi24quantum, Bhattacharjee_2021} and quantum thermal transistors \cite{PhysRevLett.116.200601}. It also offers a possible platform for testing the quantum mechanical validity of classical thermodynamic principles \cite{bera17generalized}. In addition to average thermodynamic quantities of interest, fluctuations around the mean values are crucial for measuring the performance of a quantum thermal device \cite{saryal21universal}.

Scovil and Schulz-DuBois first proposed a three-level masers as a quantum heat engine in 1959~\cite{PhysRevLett.2.262}. There have also been several studies of quantum heat engines from different thermodynamic perspectives using quantum transport. For example, Qin \textit{et al.}~\cite{PhysRevA.96.012125} investigated the role of strong system--bath coupling in a photosynthetic heat engine using a polaron master-equation approach, Ghosh \textit{et al.}~\cite{pnas.1805354115} proposed two-level masers as heat-to-work converter devices, and Lee \textit{et al.}~\cite{PhysRevE.101.022127} examined the power--efficiency trade-off in quantum thermal devices operating under nonequilibrium condition. Apart from heat engines, many other quantum thermal devices have been proposed and thoroughly studied in the last few years. These include self-contained refrigerator~\cite{PhysRevB.104.075442}, thermal transistor, switch, rectifier (diode) and other quantum machines~\cite{Ghosh2019,10.1116/5.0083192}. 
Recent studies on quantum thermal transistors have reported Floquet-engineered heat-current amplification~\cite{PhysRevE.106.024110},  heat-current switching and amplification at transient regimes~\cite{PhysRevA.103.052613}, feedback-controlled thermal management~\cite{10.1063/5.0229630}, and engineered quantum transport for heat-current modulation~\cite{PhysRevB.101.245402}. In nanoelectronics, thermal switches have been proposed based on magnetic-flux-controlled topological Josephson junctions~\cite{Sothmann_2017}, tunable multifunctional quantum thermal devices~\cite{PhysRevE.109.064146}, and stochastic thermal-fluctuation-based heat-flow control~\cite{PhysRevB.95.241401}. Quantum thermal diodes have been studied using anisotropically coupled qubits~\cite{PhysRevE.99.042121}, strongly coupled double quantum-dot circuit-QED systems~\cite{PhysRevB.99.035129}, and Coulomb-coupled fermionic quantum dots~\cite{e24121810}.
Full Counting Statistics (FCS) is a powerful tool for studying the details of current fluctuations and the heat transfer in open quantum systems and quantum thermal devices. Mandel first developed FCS for quantum optics to explain spontaneous emission of photons~\cite{Mandel:79}.  Later, Levitov and Lesovik used it to study the electron transport property through small-scale systems~\cite{10.1063/1.531672}.  Moreover, within the framework of FCS, feedback control protocols have been developed that can suppress transport noise~\cite{PhysRevB.107.125113}. In a previous study, a heat engine driven by an atomic collision has been implemented using counting statistics~\cite{Bouton2021}. 

Recently, a Floquet version of the FCS has been studied for driven thermal machines beyond the weak coupling limit and Markovian approximations~\cite{Restrepo_2018}. The FCS-based Floquet theory has also been used to describe the motion of particles in a driven quantum systems \cite{engelhardt2024,PhysRevB.102.195409}, to study fluctuations in quantum thermal machines \cite{PhysRevE.108.014137} and in quantum systems in the presence of strong system-bath coupling \cite{mahadeviya25current}. Levitov and Lesovik formulated (FCS), showing that electron transport through a conductor follows a binomial probability distribution~\cite{1993JETPL}. Cerrillo \textit{et al.} extended FCS to non-Markovian open quantum systems~\cite{PhysRevB.94.214308}. Zemach \textit{et al.} developed numerical methods for evaluating higher-order cumulants in interacting quantum systems~\cite{10.1063/5.0233876}, while Ref.~\cite{PhysRevB.109.115136} applied FCS to the efficient evaluation of transport cumulants. Notably, experimental studies of different orders of cumulants have been shown to aid in understanding electron transport in quantum dots~\cite{Flindt2009}, and also in analyzing non-Gaussian spin distributions \cite{PhysRevLett.108.183602}.

In this work, we use the Floquet master equation \cite{PhysRevE.87.012140} and the FCS method to study the operation and fluctuations in a Floquet quantum thermal transistor (FQT) \cite{PhysRevE.106.024110}. We then implement quantum control \cite{chu02cold, chakrabarti07quantum, alessandro21introduction} through the Chopped RAndom Basis (CRAB) optimization protocol \cite{PhysRevLett.106.190501, PhysRevA.84.022326}, in order to enhance the performance of the setup.  
Here we quantify fluctuations through the Fano factor, and show that one can use optimal control to significantly enhance the amplification obtained in the device. On the other hand, our results suggest that reduction of Fano factor through optimal control is associated with an increase in the base current.

\par {The paper is organized as follows: Section \ref{secII} provides a concise overview of our model. Section \ref{secIII} introduces counting field statistics techniques for calculating currents and fluctuations. Here we derive the master equation governing the dynamics of the open quantum thermodynamic system to determine the Liouvillian matrix.  We study the thermodynamic properties of FQT using FCS approaches in section \ref{SecIV}. We describe the technique to find the average energy current and fluctuation in the long-time limit in section \ref{SecIV}A, dynamic amplification factor of an FQT in section \ref{SecIV}B, the Fano factor of the setup in section \ref{SecIV}C, and the Thermodynamic Uncertainty Relation in section \ref{SecIV}D. 
We study the possibility of enhancing the performance of the FQT using optimal control in Section \ref{secV}.
Finally, we conclude in Section \ref{SecVI}, while technical details are presented in the Appendix.}\\

\section{Model}
\label{secII}

\begin{figure}
\centering
\includegraphics[width=0.85\linewidth]{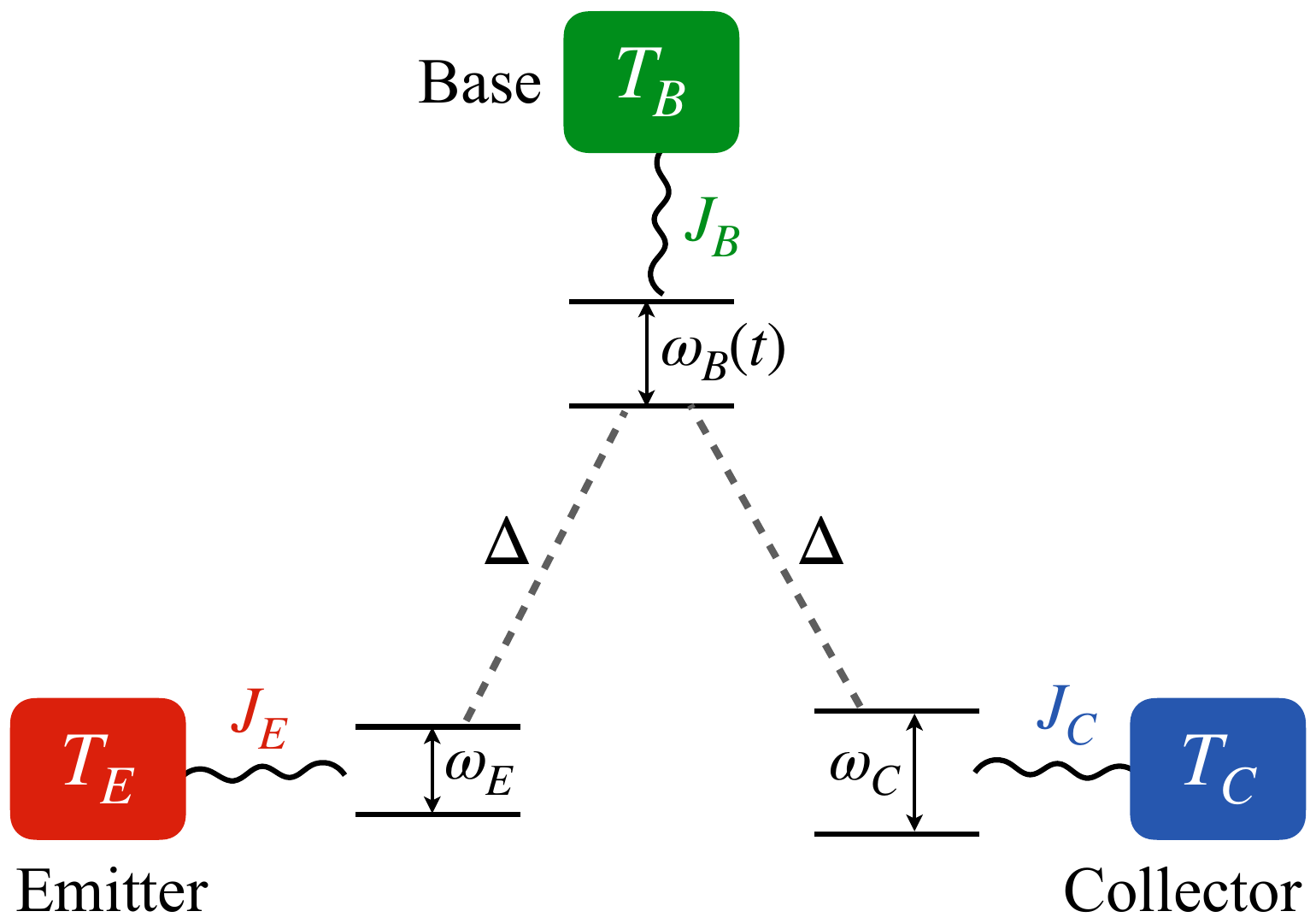}
\caption{Schematic diagram of FQT system, comprising  three terminals
qubits coupled to three thermal baths. The thermal transistor effect is achieved through periodic modulation of the frequency $\omega_B(t)$  of the base qubit. The frequencies $\omega_E$ and $\omega_C$ are the transition frequencies of the emitter and collector respectively, and $\Delta$ denotes the emitter-base and collector-base interaction strengths.}
\label{Fig_FQT}
\end{figure}

In this section, we consider a three-terminal FQT; here three two level systems (TLSs), termed Emitter ($E$), Base ($B$), and Collector ($C$), interact with three reservoirs at temperatures \(T_E\), \(T_B\), and \(T_C\), respectively, analogous to the terminals of an electronic transistor (see Fig. \ref{Fig_FQT}) \cite{PhysRevE.106.024110}.  The model Hamiltonian is given as follows 
\begin{eqnarray}
\label{model-Hamiltonian}
H(t) &=& H_{S}(t) + H_{R} + H_{I} \equiv H_0(t)+H_I\non\\
H_0(t) &=& H_S(t) + H_R, \non\\
H_S(t)&=&\frac{\hbar}{2} \omega_E \sigma_z^E + \frac{\hbar}{2} \omega_B(t) \sigma_z^B +\frac{\hbar}{2} \omega_C \sigma_z^C + \frac{\hbar}{2}\Delta(\sigma_z^E \sigma_z^B \non\\
&+& \sigma_z^B \sigma_z^C),\non\\
H_R&=& \sum_{\alpha=\{E,B,C\}} H_{R\alpha}  = H_{RE}+ H_{RB}+ H_{RC}\non\\
H_{I} &=& \sigma_{x}^{C}\otimes \mathcal{R}_{C} + \sigma_{x}^{B}\otimes \mathcal{R}_{B} + \sigma_{x}^{E}\otimes \mathcal{R}_{E}.
\end{eqnarray}
Here $H_S(t)$, $H_R$ and $H_{I}$ denotes the Hamiltonian acting on the system comprising the three TLSs, the Hamiltonian acting on the bath, and the system-bath interaction Hamiltonian, respectively, while  $H_{R\alpha}$ and $\mathcal{R}_{\alpha}$ are operators acting only the bath $\alpha = \{E, B, C\}$; \( \sigma_r^{\alpha} \) represents the Pauli matrix acting on the qubit at the terminal $\alpha$, along the axis $r = x, y, z$. 
The time-dependent TLS transition frequencies for the base terminal are  \(\omega_B(t)\), and \(\hbar \Delta/2\) represents the interaction energy between the TLSs E, B, and B, C. 
We further assume that the base terminal undergoes periodic modulation with frequency $\nu= 2\pi/\tau$, allowing $\omega_B(t)$ to take both positive and negative values while satisfying $\omega_B(t)=\omega_B(t+ \tau)$, with the time-averaged frequency $\omega_0=\frac{1}{\tau}\int_{t}^{t+\tau}dt \omega_B(t) = 0$~\cite{PhysRevA.87.013841,PhysRevE.108.014137}.  The system dynamics is controlled by product states of the 3-TLSs: 
$|1\rangle = \ket{\uparrow \uparrow \uparrow}$, $|2\rangle =  \ket{\uparrow \uparrow \downarrow}$, \( |3\rangle = \ket{\uparrow \downarrow \uparrow} \), \( |4\rangle = \ket{\uparrow \downarrow \downarrow} \), \(|5\rangle = \ket{\downarrow \uparrow \uparrow} \), \( |6\rangle = \ket{\downarrow \uparrow \downarrow} \), \(|7\rangle = \ket{\downarrow \downarrow \uparrow} \), and \(|8\rangle = \ket{\downarrow \downarrow \downarrow} \). These states are tensor product of the eigenstates $\ket{\uparrow}\; \text{and}\; \ket{\downarrow}$ of $\sigma_z^\alpha$ for each TLS in the order $\ket{i}= \ket{E,B,C}$, $i=1...8$, with corresponding eigenenergies $E_1, E_2,..., E_8$~\cite{PhysRevLett.116.200601}. The above model \eqref{model-Hamiltonian} ensures that thermal baths can flip one spin at a time. As a result, there are 12 authorized transitions, each terminal controls four transitions, as mentioned in Refs.~\cite{PhysRevLett.116.200601,PhysRevE.106.024110}. For example, the emitter induces the transitions $1 \leftrightarrow 5$, $2 \leftrightarrow 6$, $3 \leftrightarrow 7$, and $4 \leftrightarrow 8$, while the base terminal drives the transitions $1 \leftrightarrow 3$, $2 \leftrightarrow 4$, $5 \leftrightarrow 7$, and $6 \leftrightarrow 8$, and the collector one triggers the transitions $1 \leftrightarrow 2$, $3 \leftrightarrow 4$, $5 \leftrightarrow 6$, and $7 \leftrightarrow 8$. Any other transitions involving more than one spin flip are forbidden. 

We note that the model of a three-terminal quantum thermal transistor described by the Hamiltonian \eqref{model-Hamiltonian} was first proposed in Ref. \cite{PhysRevLett.116.200601} for a time independent $\omega(t)$. In this case, amplification was achieved through variation of the reservoir temperature $T_B$. On the other hand, the possibility of attaining amplification with time independent reservoir temperatures was shown in Ref. \cite{PhysRevE.106.024110}, through the modulation of the base terminal frequency $\omega_B(t)$. The asymmetry in the TLS couplings in Eq.~\eqref{model-Hamiltonian}—$EB$ and $BC$ are coupled, whereas $EC$ is not—allows the currents at $E$ and $C$ to be controlled by a small current at $B$. As shown below, the currents depend on $\nu/T_B$ and $\Delta/T_B$, allowing the amplification to be tuned via these parameters. Below we develop a FCS analysis of the FQT, and study fluctuations and optimal control in the same. As we show below, through application of optimal control and analysis of amplification and Fano factor, one can unambiguously identify a regime of ideal operation as well as a regime of worst performance, which we expect can be crucial for the incorporation of such quantum thermal transistors in near term quantum technologies \cite{campbell25roadmap}.

\section{Full Counting field statistics and Master equation}
\label{secIII} 
We assume the setup is in a product state at initial time $t = 0$, such that $\rho(0) =  \rho_S(0) \otimes \rho_R$. Here $ \rho_S(0)$ is the initial density operator of the system, while $ \rho_R = \rho_{RE} \otimes \rho_{RB} \otimes \rho_{RC}$, $\rho_{R\alpha}$ being the thermal state of the $\alpha = E, B, C$ reservoir, w.r.t. its respective Hamiltonian $H_{R_\alpha}$ and temperature ${T_{\alpha}}$. 
Let us introduce three virtual counting fields \(\chi=(\chi_E,\chi_B,\chi_C)\) to quantify the energy exchange with each thermal reservoir. Based on the two-point measurement scheme \cite{Restrepo_2018,PhysRevE.75.050102,Strasberg2021,RevModPhys.81.1665}, the statistics of the energy transferred can be evaluated using a generating function (GF), expressed as
\begin{equation}\label{eq:02}
\mathcal{G}^\chi(t) =\text{Tr}_{S}\left(\rho_S^\chi(t)\right); \quad \text{where} \quad  \rho_S^\chi(t) = \text{Tr}_{R} \left(\rho^\chi(t)\right).
\end{equation}
Here \(\text{Tr}_{R(S)}\) denotes tracing over the reservoirs (system) degrees of freedom.  The counting-field  total density matrix is given by
\begin{equation}
 {\rho}^\chi(t)= {{U}^{\chi}}(t) \rho(0) U^{-\chi}(t),\end{equation}
where \( U^{\chi}(t) = e^{-\frac{i}{\hbar}\left(\sum_{\alpha}\frac{H_{R\alpha}\chi_\alpha}{2}\right)} U(t) e^{\frac{i}{\hbar}\left(\sum_{\alpha}\frac{\chi_\alpha H_{R\alpha}}{2}\right)} \), with \(U(t)=\mathcal{T}e^{-\frac{i}{\hbar}\int_0^t H(s) ds}\), where $\mathcal{T}$ is the time ordering operator.

\vspace{-10pt} 

\subsection{Master equation}
By considering the weak coupling assumption and applying the standard Born-Markov and secular approximations, we obtain the following interaction picture master equation for the system under arbitrary modulation~(see Appendix~ \ref{Appendix-A0})
\begin{eqnarray}\label{master-eqn}
\frac{d \tilde{\rho}_S^\chi(t)}{dt} &=&  \sum_{\{\alpha=E,B,C \}}\mathcal{L_\alpha^{\chi_\alpha}}\lbrack{\tilde{\rho}_S}^\chi(t)\rbrack := \mathcal{L}^\chi \tilde{\rho}_S^\chi(t).
\end{eqnarray}
Here $\tilde{\rho}^\chi(t) = e^{-\frac{i}{\hbar}(H_S(t)+H_R)t}\rho^\chi(t) e^{\frac{i}{\hbar}(H_S(t)+H_R)t}$, and $\mathcal{L_\alpha^{\chi_\alpha}}$ are the counting field Lindblad superoperators
\begin{eqnarray}
\mathcal{L}^{\chi_\alpha}_{\alpha}  [\tilde{\rho}^\chi_{S}(t)] &=& \sum_{\{\Omega_{ij}^{\alpha}\}}\bigg[G_\alpha(\Omega_{ij}^\alpha)\mathcal{D}(\mathcal{A}_\alpha)[\tilde{\rho}^\chi_S] \nonumber\\&+& G_\alpha(-\Omega_{ij}^\alpha)\mathcal{D}(\mathcal{A}^{\dagger}_\alpha)[\tilde{\rho}^\chi_S])\bigg];\quad \alpha=\{E,C\},\\
\mathcal{L}^{\chi_B}_{B}  [\tilde{\rho}^\chi_{S}(t)] &=&\sum_{q \in \mathbb{Z},\{\Omega_{ij}^B \}}P_q\bigg[ G_B(\Omega_{ij}^B+q\nu) \mathcal{D}(\mathcal{A}_\mathcal{B})[\tilde{\rho}^\chi_S]\nonumber \\&+& G_B(-\Omega_{ij}^B-q\nu) \mathcal{D}(\mathcal{A}^{\dagger}_{\mathcal{B}})[\tilde{\rho}^\chi_S]\bigg],
\end{eqnarray}
in terms of the counting field modified dissipator 
\begin{eqnarray}
\mathcal{D}(\mathcal{A}_\alpha)[\tilde{\rho}^\chi_S] = e^{-i \Omega_{ij}^\alpha \chi_\alpha } \mathcal{A}_\alpha \tilde{\rho}^\chi_S(t) \mathcal{A}_\alpha^\dagger  - \frac{1}{2} \{\mathcal{A}_\alpha^\dagger \mathcal{A}_\alpha,\tilde{\rho}^\chi_S(t)\} \non\\
\end{eqnarray}

The operator $\mathcal{A}_\alpha$ takes on the form $|i\rangle \langle j|$ ($i \neq j$; $i, j = 1, 2, . . . , 8)$ and causes the 12 transitions mentioned above (see Sec. \ref{secII}). 
Here, the transition frequency between the energy levels $\ket{i}$ and $\ket{j}$ mediated by $\alpha$-th bath is given by \(  \Omega_{ij}^\alpha = \left(E_i-E_j\right)_{\alpha}/\hbar > 0\), where the subscript $\alpha$ on the r.h.s signifies the constraint on the  transitions allowed for the $\alpha$-th bath (see Sec. \ref{secII}); 
\( q \in \mathbb{Z} \) denote the distinct Floquet modes with frequencies \( \Omega_{ij}^B + q \nu \) and amplitudes $P_q = P_{-q}$, such that \( \sum_{q \in \mathbb{Z}} P_q = 1 \) \cite{PhysRevE.87.012140, PhysRevE.94.062109}. \( G_\alpha(\Omega_{ij}^\alpha) = G_0(\Omega_{ij}^\alpha)[1 + \bar{n}_\alpha(\Omega_{ij}^\alpha) ] \) is the spectral function of  the $\alpha$-th Bosonic bath 
with average thermal occupancy \( \bar{n}_\alpha(\Omega_{ij}^\alpha) = (e^{{\hbar\Omega_{ij}^\alpha}/{k_B T_\alpha}} - 1)^{-1} \)  at  frequency \( \Omega_{ij}^\alpha \) and temperature \( T_\alpha \). We consider Ohmic baths, so that \( G_{0}(\Omega_{ij}^\alpha) = \kappa \Omega_{ij}^\alpha \), where the constant \( \kappa >0 \) is the same for all three reservoirs.

In the steady state, we have $\dot{\tilde{\rho}}_{ss}^\chi=0$, i.e., the populations are diagonal in the energy eigenbasis $\{|j\rangle\}$. As a result, Eq.~\eqref{master-eqn} reduces to 
\begin{eqnarray}
\dot{\tilde{\rho}}^{\chi}_{11} &=0= \Gamma^{\chi_E}_{51} + \Gamma^{\chi_B}_{31} + \Gamma^{\chi_C}_{21} ; \nonumber\\
\dot{\tilde{\rho}}^{\chi}_{22} &=0= \Gamma^{\chi_E}_{62} + \Gamma^{\chi_B}_{42} + \Gamma^{\chi_C}_{12} ; \nonumber\\
\dot{\tilde{\rho}}^{\chi}_{33} &=0= \Gamma^{\chi_E}_{73} + \Gamma^{\chi_B}_{13} + \Gamma^{\chi_C}_{43} ; \nonumber\\
\dot{\tilde{\rho}}^{\chi}_{44} &=0= \Gamma^{\chi_E}_{84} + \Gamma^{\chi_B}_{24} + \Gamma^{\chi_C}_{34} ;\nonumber \\
\dot{\tilde{\rho}}^{\chi}_{55} &=0= \Gamma^{\chi_E}_{15} + \Gamma^{\chi_B}_{75} + \Gamma^{\chi_C}_{65} ; \nonumber\\
\dot{\tilde{\rho}}^{\chi}_{66} &=0= \Gamma^{\chi_E}_{26} + \Gamma^{\chi_B}_{86} + \Gamma^{\chi_C}_{56} ; \nonumber\\
\dot{\tilde{\rho}}^{\chi}_{77} &=0= \Gamma^{\chi_E}_{37} + \Gamma^{\chi_B}_{57} + \Gamma^{\chi_C}_{87} ; \nonumber\\
\dot{\tilde{\rho}}^{\chi}_{88} &=0= \Gamma^{\chi_E}_{48} + \Gamma^{\chi_B}_{68} + \Gamma^{\chi_C}_{78} ; 
\end{eqnarray}
Here, the net transition rates $\Gamma^{\chi_\alpha}_{ij}$, are determined by the following expressions:
\( \Gamma^{\chi_\alpha}_{ij}= G_{\alpha}(\Omega_{ij}^\alpha) \tilde{\rho}_{ii}^\chi e^{- i \Omega_{ij}^\alpha\chi_\alpha} - G_{\alpha}(-\Omega_{ij}^\alpha) \tilde{\rho}_{jj}^\chi, \quad \alpha = \{E, C\}, \) and
\(\Gamma^{\chi_B}_{ij}=\sum_{q} \Gamma^{\chi_B}_{ij,q} \),
where, $ \Gamma^{\chi_B}_{ij,q} = P_{q} [G_{B}(\Omega_{ij}^B + q\nu)e^{-i(\Omega_{ij}^B+ q \nu)\chi_B} \tilde{\rho}_{ii}^\chi - G_{B}(-\Omega_{ij}^B - q\nu) \tilde{\rho}_{jj}^\chi]$,~ and~ ${\Gamma^{\chi_\alpha}_{ij}}/{\Gamma^{\chi_\alpha}_{ji}} =e^{-i \Omega_{ij}^\alpha\chi_\alpha} $.
Under the condition \( \omega_{E} =\omega_{C} = 0 \), and $\Delta \gg \omega_{B}(t)$ $\forall~t$~\cite{PhysRevLett.116.200601},  states $|1\rangle$ and $|8\rangle$ become degenerate and assigned as $|I\rangle$, similarly, states $|2\rangle$ and $|7\rangle$ are denoted as $|II\rangle$, states $|3\rangle$ and $|6\rangle$ are termed as $|III\rangle$, and states $|4\rangle$ and $|5\rangle$ are termed $|IV\rangle$, (cf. Refs.~\cite{PhysRevLett.116.200601,PhysRevE.106.024110}). Consequently, the new density matrix is characterized by four elements: $\tilde{\rho}_{I}^\chi = \tilde{\rho}_{11}^\chi + \tilde{\rho}_{88}^\chi,~\tilde{\rho}_{II}^\chi = \tilde{\rho}_{22}^\chi + \tilde{\rho}_{77}^\chi,~\tilde{\rho}_{III}^\chi = \tilde{\rho}_{33}^\chi  +\tilde{\rho}_{66}^\chi,$ and $\tilde{\rho}_{IV}^\chi = \tilde{\rho}_{44}^\chi + \tilde{\rho}_{55}^\chi$ and the dynamics of \( \tilde{\rho}_S^\chi(t) \) can be represented by a $4\times4$ Liouvillian matrix \(\mathcal{L}^\chi\) acting on the vectorized \( \tilde{\rho}_S^\chi(t) = [\tilde{\rho}_{I}^\chi(t), \tilde{\rho}_{II}^\chi(t), \tilde{\rho}_{III}^\chi(t), \tilde{\rho}_{IV}^\chi(t)]^T \)  (see Appendix~\ref{Appendix-2}). Considering up to the first two harmonics ($q = 0, \pm 1$ and $q' = \pm 1 $), matrix \(\mathcal{L^\chi}\), takes the following form 
\begin{widetext}
\begin{equation}\label{Matrix}
\mathcal{L}^\chi = \left(
\begin{array}{cccc}
\mathcal{Q} & \Delta e^{-\frac{\hbar\Delta}{k_BT_C}+i \Delta \chi_C} & 2 \Delta R(\nu) e^{-\frac{2\hbar \Delta}{k_B T_B}+2 i \Delta \chi_B} & \Delta e^{-\frac{\hbar\Delta}{k_B T_E}+i \Delta \chi_E} \\
\Delta e^{-i \Delta \chi_C} & -\mathcal{B}-\Delta e^{-\frac{\hbar\Delta}{k_BT_C}}-\Delta & \Delta e^{-\frac{\hbar\Delta}{k_BT_E}+i \Delta \chi_E} & \mathcal{B} \\
2 \Delta R(0) e^{-2 i \Delta \chi_B} & \Delta e^{-i \Delta \chi_E} & -2 \Delta R(\nu) e^{-\frac{2\hbar \Delta}{k_BT_B}}-\Delta \left(e^{-\frac{\hbar\Delta}{k_BT_C}}+e^{-\frac{\hbar\Delta}{k_BT_E}}\right) & \Delta e^{-i \Delta \chi_C} \\
\Delta e^{-i \Delta \chi_E} & \mathcal{B} & \Delta e^{-\frac{\hbar\Delta}{k_BT_C}+i \Delta \chi_C} & -\Delta -\mathcal{F}-\Delta e^{-\frac{\hbar\Delta}{k_BT_E}} \\
\end{array}
\right),
\end{equation}
\end{widetext}
where the system parameters are chosen as \(\{T_E, T_B, T_C\} \ll \hbar\Delta/k_B\), and we define  
\[\mathcal{F}=  \nu P_1 \coth \left(\frac{\hbar\nu}{2k_B T_B}\right), \quad \mathcal{B}=P_0 k_BT_B/\hbar +\mathcal{F},\]
\[ R(\nu) = P_0 + P_1 \left(\frac{\nu}{2 \Delta} + 1\right) e^{-\frac{\hbar\nu}{k_BT_B}} + P_1 \left(1 - \frac{\nu}{2 \Delta}\right) e^{\frac{\hbar\nu}{k_BT_B}},\]  and $\mathcal{Q}= -2\Delta (1+R(0))$, respectively.  The counting fields do not affect the relaxation dynamics of the FQT, as the decay rates depend on the real part of the Liouvillian eigenvalues at zero counting fields. As shown in Ref.~\cite{PhysRevE.106.024110}, the ratio $\Gamma/\Delta$ increases approximately linearly with $k_B T_B/(\hbar\Delta)$ and remains of the order of $10^{-3}$ i.e., $\Gamma/\Delta \ll 1$, in the intermediate-base temperature  regime ($T_B<T_E$) where the FQT operates most efficiently. Since $\Gamma/\Delta \ll 1$, this confirms the validity of the Born--Markov approximation. The same holds in the presence of modulation in the FQT, as shown in Ref.~\cite{PhysRevE.106.024110}.

\section{Thermodynamic Characteristics of Transistors}\label{SecIV}
\subsection{Generating function, mean currents and fluctuations}
\label{SecIVA}The formal solution of the counting field-dependent master equation~\cite{Blanter2000} can be expressed as
\begin{equation}
\tilde{\rho}_S^\chi(t) = e^{\mathcal{L}^\chi t} \tilde{\rho}_S(0)\end{equation}
One can define the cumulant GF (CGF) from Eq.~\eqref{eq:02} as $ \mathcal{C}^\chi(t)\equiv \ln{\mathcal{G}^{\chi}(t)}=  \ln \text{Tr}[\tilde{\rho}_S^\chi(t)]$. 
\begin{figure*}\centering
\includegraphics[width=1\linewidth]{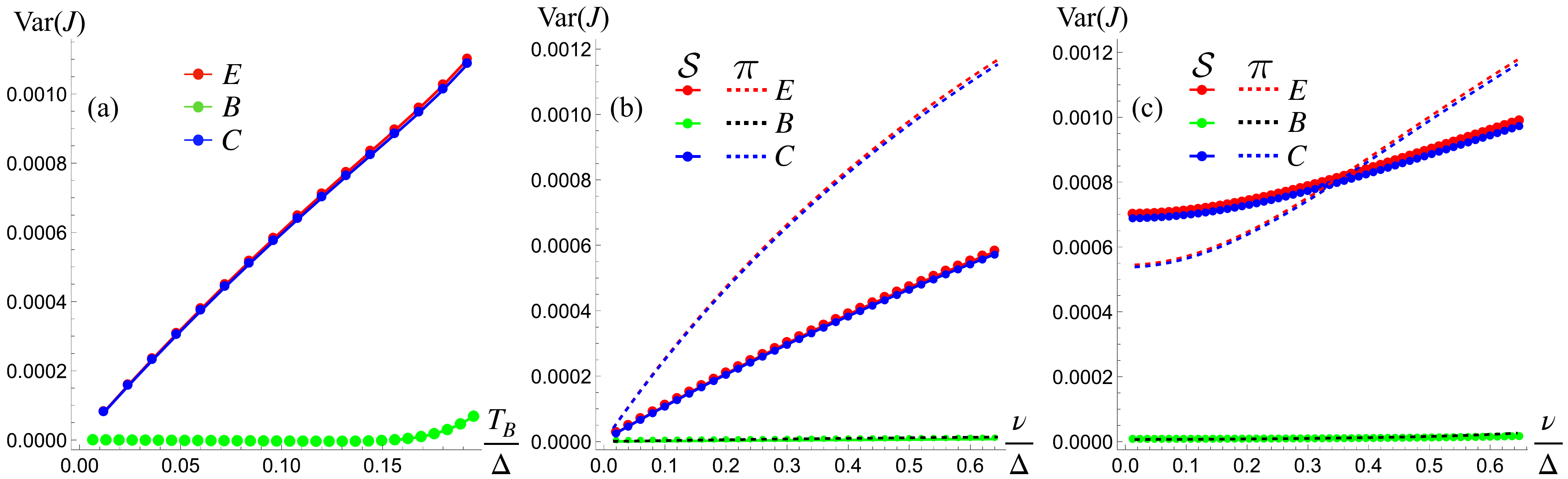}
\caption{Figure showing current variance Var($J$) as a function of (a) $T_B/\Delta$ for the unmodulated case,(b) $\nu/\Delta$ for sinusoidal ($\mathcal{S}$) and $\pi$-flip modulation  for $T_B \rightarrow 0$ and \(\lambda = 0.8\), and (c) $\nu/\Delta$ under sinusoidal and $\pi$-flip modulation for finite temperatures $T_B = 0.118\Delta$ and \(\lambda = 0.8\).  Here $\omega_C= \omega_0=\omega_E=0$, $T_E=0.2\Delta$, $T_C=0.02 \Delta$ and $\Delta = 1$.} 
 \label{F002}
\end{figure*}
\begin{figure*}\centering
\includegraphics[width=1\linewidth]{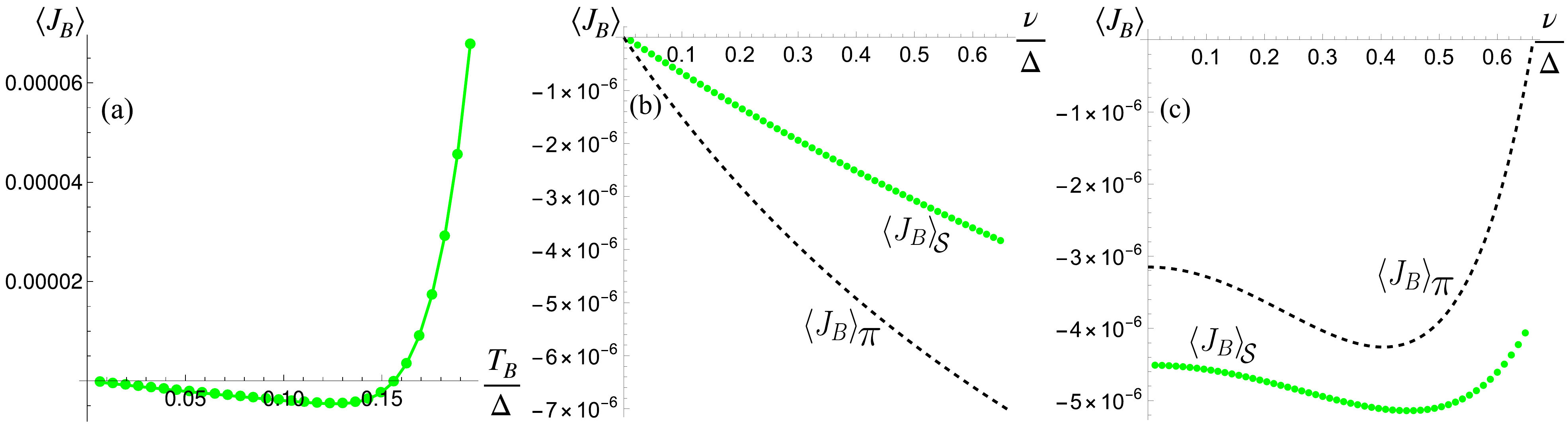}
\caption{The plot shows the base current as a function of (a) $T_B/\Delta$ for unmodulated case, (b) $\nu/\Delta$ for sinusoidal ($\mathcal{S}$) and $\pi$-flip modulation for $T_B \rightarrow 0$ and \(\lambda = 0.8\), and (c) $\nu/\Delta$ for finite temperatures $T_B = 0.118\Delta$ and \(\lambda = 0.8\). Other parameters are the same as in Fig.~\ref{F002} (see also Ref. \cite{PhysRevE.106.024110}). }
 \label{Figs_JB}
\end{figure*}
 Numerically one finds that real parts of the eigenvalues of the matrix $\mathcal{L}^\chi(t)$ are negative and the long-time behavior of $\tilde{\rho}_S^\chi(t)$ is primarily influenced by the largest eigenvalue $\tilde{\lambda}^\chi \to 0^{-}$ of $\mathcal{L}^\chi(t)$. Therefore in the steady state, we can approximate the CGF as follows ~\cite{Sornette2006,RevModPhys.81.1665,Schaller2014} {\textbf{\begin{equation}\lim_{{t \to \infty}}\mathcal{C^\chi}(t) \approx {\tilde{\lambda}}^\chi t 
\end{equation}}and calculate the average heat current as well as the higher order cumulants~\cite{RevModPhys.81.1665}; the $k$-th order cumulant of the heat current in the steady state is given by~\cite{PhysRevResearch.5.023155}:
\begin{equation}\langle\langle J_\alpha^k \rangle\rangle = \lim_{{t \to \infty}}\frac{d}{dt}\frac{{d^k }}{{d(i\chi_\alpha)^k}}\mathcal{C}^\chi( t) \Big|_{\chi=0} = \frac{{d^k {\tilde{\lambda}}^\chi}}{{d(i\chi_\alpha)^k}}\Big|_{\chi=0}. \end{equation}
The expression for the mean heat current can be written as follows (see details in the Appendix~\ref{J-alpha})
\begin{equation}
\langle J_\alpha\rangle = \frac{d\tilde{\lambda}^\chi}{d(i\chi_\alpha)} \Big|_{\chi=0}= -\frac{A_0^{(1)}}{A_1^{(0)}}.
\end{equation}
Here, $A_n$ and $A_n^{(k)}$  denote the expansion coefficients of the polynomial \(P^{\chi}(z)=\det(zI-\mathcal{L}^{\chi})\)~defined as
\begin{equation}
A_n=\frac{1}{n!}\frac{\partial^n P^{\chi}(z)}{\partial z^n}\Big|_{z=0},\qquad
A_n^{(k)}=\frac{\partial^k A_n}{\partial (i\chi)^k}\Big|_{\chi=0}.
\end{equation}
We note that $\langle J_\alpha\rangle > 0$ indicates that the system is absorbing heat from the $\alpha$-th bath, whereas $\langle J_\alpha\rangle < 0$ denotes that heat is flowing into the bath. We consider the regime where $e^{- \hbar\Delta/k_B T_C} \ll e^{-\hbar\Delta/k_B T_E} \ll 1$. 
Finally, 
we obtain the following average heat current expressions [see Appendix~\ref{Appendix-A}: Eqs. \eqref{dot1}-\eqref{dot3}]:}
\begin{equation}\label{EQ.1}
{\langle J_E\rangle}\simeq\frac{ \hbar\Delta ^3}{\mathcal{Y}_1} \mathcal{B} e^{-\frac{\hbar\Delta}{k_BT_E}}\left[ e^{-\frac{\hbar\Delta}{k_BT_E}}( 4R(0)+1)+2( R(0)+1)\right],\end{equation}
where $\mathcal{Y}_1=2( R(0)+1)\left[-\mathcal{B}^2+(\mathcal{B}+\Delta)(\mathcal{F}+\Delta)\right]$.
\begin{equation}\label{EQ.2}
\langle J_B \rangle \simeq \frac{4\hbar \Delta^3}{\mathcal{Y}_1} \left[ \mathcal{B} \Psi + e^{-\frac{2\hbar\Delta}{k_B T_B}} R(\nu) (\mathcal{F} + 2\Delta) \right],
\end{equation}  
where  
\[
\Psi = 3R(\nu)e^{-\frac{2\hbar\Delta}{k_B T_B}}  -  R(0)e^{-\frac{2\hbar\Delta}{k_B T_E}}.
\]
\begin{equation}\label{EQ.3}
{\langle J_C\rangle}\simeq -\frac{ \hbar\Delta ^3}{\mathcal{Y}_1} \mathcal{B}e^{-\frac{\hbar\Delta}{k_BT_E}}\left[ (1+R(0))+e^{-\frac{\hbar\Delta}{k_BT_E}}\right]
\end{equation}
We analyze the behavior of average current and noise power in a thermal transistor under sinusoidal and $\pi$-modulations. In case of sinusoidal modulation, we consider 
\begin{equation}
\omega_B(t) = \lambda \nu \sin(\nu t),   
\label{eqsin}
\end{equation} 
with \( 0 \leq \lambda \leq 1 \). In the limit of weak modulation, one can restrict the analysis to the first two harmonics  \( q = 0, \pm1 \), given by~\cite{PhysRevE.87.012140}
\begin{equation} P_0 = 1 - \frac{\lambda^2}{2}, \quad P_{\pm1} = \frac{\lambda^2}{4}. \end{equation} 
Similarly, the {$\pi$-flip} modulation is expressed as 
\begin{equation} 
\omega_B(t) = \omega_0 + \sum_{q \in \mathbb{Z}} \pi[\delta(t - \tau(q + 1/4)) - \delta(t - \tau(q + 3/4))]
\label{eqpi}
\end{equation}
which results in only two primary harmonics $q = \pm 1$, with~\cite{PhysRevE.87.012140}.
\begin{equation}
P_0 = 0 \quad \text{and} \quad P_{\pm1} \approx \left(\frac{2}{\pi}\right)^2.  
\end{equation}
For the pulses Eqs. \eqref{eqsin} and \eqref{eqpi} considered above, one can average out over a few modulation time periods $\tau$, such that in the limit of weak system-bath coupling, and bath correlation time $\ll \tau$, the Born, Markov and secular approximations remain valid \cite{PhysRevE.87.012140, takashi23floquet}. This in turn results in the Markovian dynamics considered here. We note that this formalism has been used extensively for designing continuously modulated quantum heat engines \cite{PhysRevE.87.012140, alicki2012periodicallydrivenquantumopen, PhysRevE.108.014137}, quantum thermometers \cite{mukherjee19enhanced} and FQT \cite{PhysRevE.106.024110}, and also has been studied experimentally \cite{chen25engineering}.
  Using FCS, the cycle-averaged power is obtained by  the first cumulant (mean) of the work distribution~\cite{Silaev2014,Strasberg2021} per unit cycle. As shown in Eqs.~\eqref{EQ.1}--\eqref{EQ.3} and  for the numerical parameters used  in Fig.~\ref{F002}, we find that the average total heat current  $\langle J_E \rangle + \langle J_B \rangle + \langle J_C \rangle \approx 0$, 
 which indicates that the system follows  the energy conservation law in the steady state [See Fig.~\ref{F01} in Appendix~\ref{Appendix-A}]. Consequently, the cycle-averaged power associated with the periodic driving is zero, which means no net power is exchanged between the external modulation and the system.
Now, let us consider the second cumulant, which denotes the width of the current distribution. One can define~\cite{PhysRevResearch.5.023155} the current fluctuation through \(\text{Var} {(J_\alpha)} \) for $\alpha$-th bath as:
\begin{equation}\label{EQ.var}
\text{Var} ({J}_\alpha) = \frac{d^2\tilde{\lambda}^\chi}{d(i\chi_\alpha)^2} \Big|_{\chi=0}=-\frac{ A_0^{(2)} + 2 A_1^{(1)}\langle J_\alpha\rangle + 2A_2^{(0)}\langle J_\alpha\rangle^2}{A_1^{(0)}}
\end{equation}
As shown in Eq. \eqref{EQ.var}, the variance depends on the mean currents, which in turn are given by Eqs. \eqref{EQ.1} - \eqref{EQ.3}. For numerical plots, we set $\hbar=k_B=1$. In Fig.~\ref{F002}(a), we show that base current noise significantly increases with higher $T_B$, indicating that as $T_B \rightarrow T_E$, the Born-Markov approximation breaks down, and our model ceases to function as a viable heat modulation device (see Fig. \ref{Figs_JB}). The above effect arises since the Born–Markov approximation remains valid as long as the inverse of the relevant frequency difference in the problem (taken to be $\Delta \sim \mathcal{O}(1)$) is small compared to the relaxation time of the system. As shown in Ref.~\cite{PhysRevLett.116.200601,PhysRevE.106.024110}, the decay rates of the energy level increase with temperature, thus resulting in the above approximation to break down in the high-$T_B$ regime.  
In the presence of modulation, owing to different forms of the Floquet amplitudes $P_q$, the current noise amplitude (Cf. Fig.~\ref{F002}(b)), as well as the magnitude of the current (Fig.~\ref{Figs_JB}b), are higher for the $\pi$-flip modulation than that for a sinusoidal modulation in the limit of $T_B \rightarrow 0$. The corresponding forms for intermediate values of $T_B$ are shown in Figs \ref{F002}c and ~\ref{Figs_JB}c.

\begin{figure}
\centering
\includegraphics[width=1\linewidth]{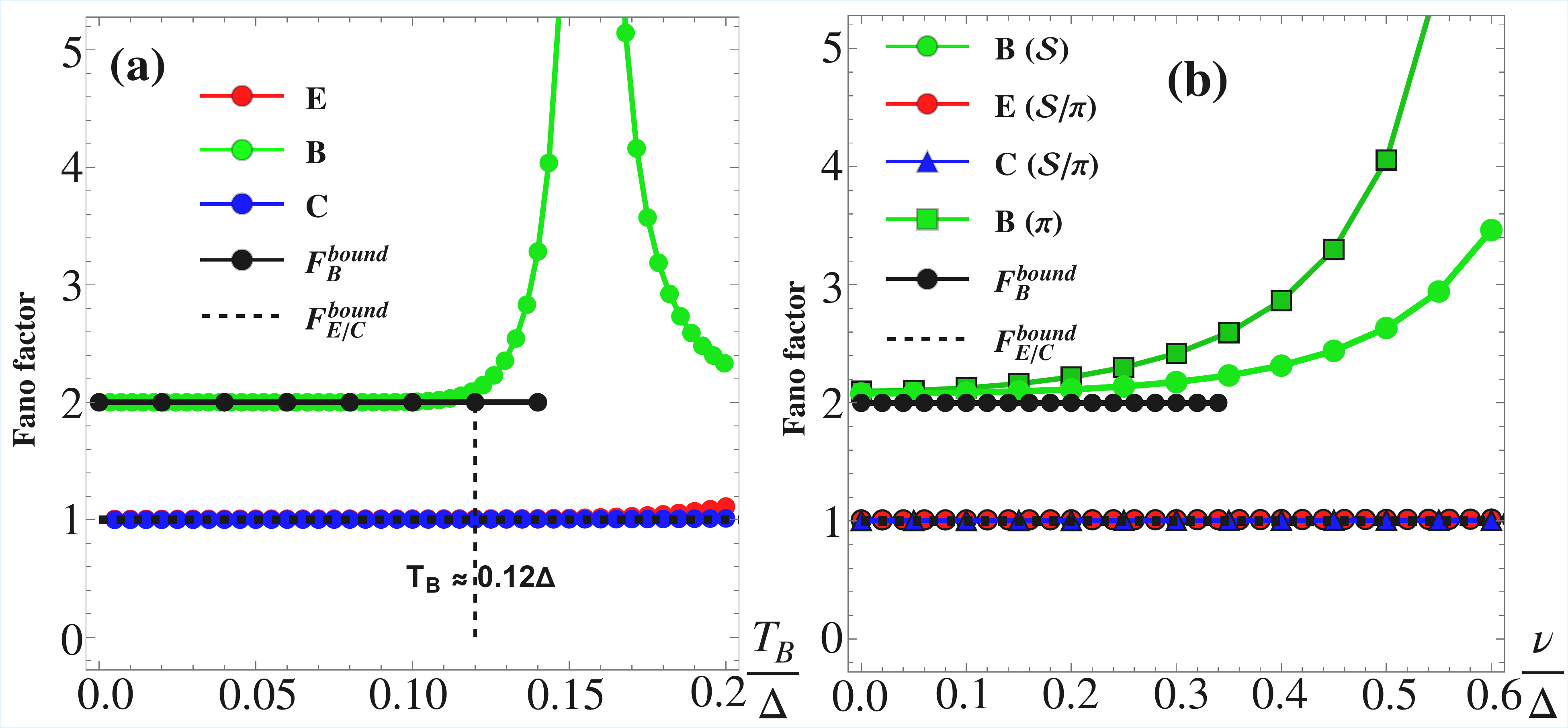}
\caption{(a) The plot shows the Fano factor as a function of $T_B/\Delta$ for the unmodulated case. The parameters are the same as those used in Fig.~\ref{F002}(a). Plot (b) shows the behavior of the Fano factor versus $\nu/\Delta$  for sinusoidal modulation and the $\pi$-flip modulation cases. The parameters are the same as those used in Fig.~\ref{F002}(c). For both plots, the horizontal  lines (black solid  for $F_B$ and  black thick dashed for $F_{E/C}$)  represent the lower bounds of the Fano factor, given by the r.h.s. of Eq. \eqref{f10}, which satisfy  $F_{B}^{\text{bound}} = 2$ and $F_{E/C}^{\text{bound}} = 1$.} \label{F10}
\end{figure}

\subsection{Amplification factor} 
In a thermal transistor, one gets a significant change in the emitter and collector terminal currents, through a small change in the base current. 
The operation of a thermal transistor can be quantified through the dynamical amplification factors, defined as~\cite{PhysRevE.106.024110} 
\begin{equation}\label{EQ:5}\beta_+^{x}=\frac{\partial{\langle J_C\rangle }}{\partial{\langle J_B\rangle}}, ~~~~\beta_-^{x}=\frac{\partial{\langle J_E\rangle}}{\partial{\langle J_B\rangle}}.
\end{equation}
Here $x$ is the tuning parameter, which corresponds to either the base temperature $T_B$, or to the modulation frequency $\nu$, either of which can be tuned to achieve amplification.
An amplification is achieved when the amplification factor $|\beta_\pm^{x}| > 1$.
As discussed in detail in Ref.~\cite{PhysRevE.106.024110}, and also presented in Appendix \ref{Appendix-A}, $\langle J_B\rangle$ attains its minimum at $T_B\simeq0.125\Delta $, which also results in the  divergence of $\beta_{\pm}^{x}$ at around this temperature. In this regime, an infinitely  small change in $\langle J_B\rangle$  makes pronounced variations in $\langle J_{E/C}\rangle$, indicating thermal transistor responds very sensitively and amplifies significantly (see Eq. ~\eqref{beta-app}).  For $T_B \ll T_E$, we obtain analytical approximations of   \(\beta_{\pm}^{x}\)  for the two types of modulation cases:
\begin{equation}\label{s-Pi}
 \beta=~ 
\begin{cases} 
\begin{aligned}
&\text{Sinusoidal modulation:} \\
&\quad \beta_-^{\nu} \simeq - e^{\hbar \Delta / k_B T_E} - 1.25, \\ 
&\quad \beta_+^{\nu} \simeq  e^{\hbar \Delta / k_B T_E} + 0.25, \\[1em]
&\text{$\pi$-flip modulation:} \\
&\quad \beta_-^{\nu} \simeq -\frac{2(8 + \pi^2) e^{\hbar \Delta / k_B T_E} + 32 + \pi^2}{32}, \\ 
&\quad \beta_+^{\nu} \simeq \frac{2(8 + \pi^2) e^{\hbar \Delta / k_B T_E} + \pi^2}{32}.
\end{aligned}
\end{cases}
\end{equation}

As shown in Appendix~\ref{Appendix-A}, $\pi$-flip modulation results in higher amplification as  compared to that obtained for sinusoidal modulation; this can be attributed to the different forms of pre-factors to the exponential factor $e^{\hbar \Delta/k_B T_E}$ for the two modulation schemes (see Eq. \eqref{s-Pi}).

\subsection{Fano Factor} The Fano factor $F_\alpha$, defined as
\begin{equation} \label{Eq:fano} F_\alpha =\frac{\text{Var}({ J_\alpha})}{|{\langle J_\alpha\rangle}|},
\end{equation}
quantifies how precisely one can control the heat current relative to the thermal noise of the device. 
Furthermore, Fano factor can be experimentally measured from the counting statistics \cite{Wagner2019,PhysRevLett.108.183602}, thus making its study and optimization highly relevant for developing heat modulation devices. 
We compare the Fano factor for the unmodulated and modulated scenarios, respectively, in Figs.~\ref{F10}(a) and~\ref{F10}(b). In the device, the base current is significantly smaller in magnitude than the emitter and collector currents. Since the Fano factor is defined as the ratio of the current variance to its mean value, a small average current enhances the relative contribution of fluctuations. Consequently,  as shown in Fig.~\ref{F10} , the base terminal exhibits a larger Fano factor ($F_B \gtrsim 2$), indicating pronounced super-Poissonian noise and reduced transport precision. In contrast, the emitter and collector terminals satisfy $F_{E/C} \approx 1$, consistent with nearly Poissonian statistics and comparatively stable transport. The Fano factor thus provides a quantitative measure of current stability and noise amplification in multi-terminal thermal devices, and also offers a convenient way to identify classical and nonclassical behavior in thermal currents~\cite{PhysRevE.97.052145}. The base terminal ($F_B \gtrsim 2$) is not anomalous, but rather a direct consequence of operating the device in a low-current control regime, where fluctuations dominate over the mean heat flow.

As mentioned earlier, here we tune the currents $\langle J_{\alpha}\rangle$ by tuning $\nu/\Delta$. Consequently, as shown in Fig.~\ref{F10}b, for a fixed $T_B/ \Delta~(= 0.118$, see Fig. \ref{F002}c), modulating the base frequency changes the Fano factor at the base terminal as well. The increase in $F_B$ with $\nu/\Delta$ is smaller for sinusoidal modulation than for $\pi$-modulation. 

Recently, Guarnieri et. al.~\cite{PhysRevLett.133.070405} have shown that the Fano factor for a quantum system—irrespective of whether it is harmonic or spin—in the presence of external time-dependent driving is given by $F_\alpha \geq \hbar \langle \omega \rangle \coth \left( \beta \hbar \langle \omega \rangle / 2 \right)$, where, $\langle .\rangle$ denotes the average with respect to the pseudo-mode distribution. Following this result and based on the numerical fitting for both the unmodulated as well as the modulated cases, we make an ansatz, in the spirit of the above expression, to estimate the Fano factor for the $\alpha$-th current which is given 
\begin{eqnarray}\label{f10}
F_\alpha & \geq\hbar\Omega_\alpha\coth{\left(\frac{\hbar\Omega_\alpha\langle \sigma \rangle}{2 k_B |\langle J_\alpha\rangle|}\right)}
\end{eqnarray}  
where the average entropy production rate \( \langle \sigma \rangle \) for the transistor is $\langle \sigma \rangle =  -\sum_{\alpha= E,B,C} \langle J_\alpha \rangle/T_\alpha $, and we identify $\Omega_B = 2\Delta$ and $\Omega_{E, C} = \Delta$.  In the steady state, the total energy (heat) current will be conserved i.e. $\langle J_E\rangle+\langle J_B\rangle+\langle J_C\rangle=0$.  For $|J_{E/C}| \gg |J_B|$, we have $J_E \approx -J_C$, which eventually results in 
\begin{eqnarray}
	F_{\alpha} &\gtrsim& \hbar \Omega_{\alpha} \coth\!\left[\frac{\hbar \Omega_{\alpha}}{2 k_B}\left(\frac{1}{T_C} - \frac{1}{T_E}\right)\right] \nonumber \\
	&\gtrsim& \hbar \Omega_{B} \coth\!\left[\frac{\hbar \Omega_{B} J_E}{2 k_B J_B}\left(\frac{1}{T_C} - \frac{1}{T_E}\right)\right].
	\label{FanoBound}
\end{eqnarray}
Here we have assumed $|J_{E}|/T_E,~ |J_C|/T_C \gg |J_B|/T_B$. For $J_E/J_B \to \infty$, we have $F_B \gtrapprox \hbar\Omega_B$. For the given parameters in Fig.~\ref{F002}, we observe that the Fano factor for the emitter/collector case remains constant across the wide range of base temperature variations, in good agreement with Eq. \eqref{FanoBound}. However, for the base terminal exhibiting periodic modulation, $F_B$ saturates its lower bound only
for small values of $T_B/\Delta$ and modulation frequency $\nu/\Delta$.  Vanishing $\langle J_B \rangle$ at $T_B /\Delta \approx 0.16$ and at $\nu/\Delta \approx 0.66$  (see Eq. \eqref{EQ.2} and Fig. \ref{Figs_JB}) result in  diverging $F_B$ (see Fig. \ref{F10}).

\subsection{Thermodynamic Uncertainty Relation}
In this section, we quantify the performance of the FQT through the Thermodynamic Uncertainty Relation (TUR), which is evaluated through the product of the total entropy production and the noise-to-signal ratio. Notably, TUR has been used extensively in studies involving fluctuations in systems out of equilibrium \cite{koyuk22thermodynamic, Horowitz2020, agarwalla18assessing} and in quantum engines \cite{PhysRevE.108.014137, singh23thermodynamic}, and more recently, in three-terminal  systems \cite{zhang2025thermodynamic}.
Let us start with the definition of the entropy production rate
\begin{equation}
    \langle\sigma\rangle=-\sum_{\alpha=E, B, C} \frac{\langle J_\alpha\rangle}{T_\alpha}.
\end{equation}
At the steady state one can rewrite the entropy production rate as follows
\begin{eqnarray}
    \langle\sigma\rangle=\langle J_E\rangle\Bigg(\frac{1}{T_B}-\frac{1}{T_E}\Bigg)-\langle J_C\rangle\Bigg(\frac{1}{T_C}-\frac{1}{T_B}\Bigg).
\end{eqnarray}
In case of FQT, we assume $T_E>T_B>T_C$. This leads to $\langle J_E\rangle>0$ and $\langle J_C\rangle<0$, resulting the positive entropy production rate ($\sigma>0$). For transistor operation, the emitter and collector heat currents must dominate over the base current: $|\langle J_E\rangle|,|\langle J_C\rangle|>>|\langle J_B\rangle|$. In the low temperature regime $\{T_E,T_B,T_C \}<< \hbar\Delta/k_B$, we have shown that $\langle J_E\rangle\approx -\langle J_C\rangle$ (see appendix~\ref{Appendix-A}). With this, the entropy production rate can be rewritten as
\begin{equation}\label{approx_sigma_eq}
    \langle\sigma\rangle \approx |\langle J_E\rangle|\Bigg(\frac{1}{T_C}-\frac{1}{T_E} \Bigg); \quad \text{or}\quad \langle\sigma\rangle \approx |\langle J_C\rangle|\Bigg(\frac{1}{T_C}-\frac{1}{T_E} \Bigg).
\end{equation}
The resultant TUR ratio is given by
\begin{eqnarray}\label{TUR}
Q_{J_\alpha}=\frac{\text{Var}(J_\alpha)\langle\sigma\rangle}{|\langle J_\alpha \rangle|^2 k_B}=F_\alpha\frac{\langle\sigma\rangle}{  |\langle J_\alpha \rangle|  k_B},\; \alpha=E,B,C,
\end{eqnarray}
 
\begin{figure}[!h]
\centering \includegraphics[width=1\linewidth]{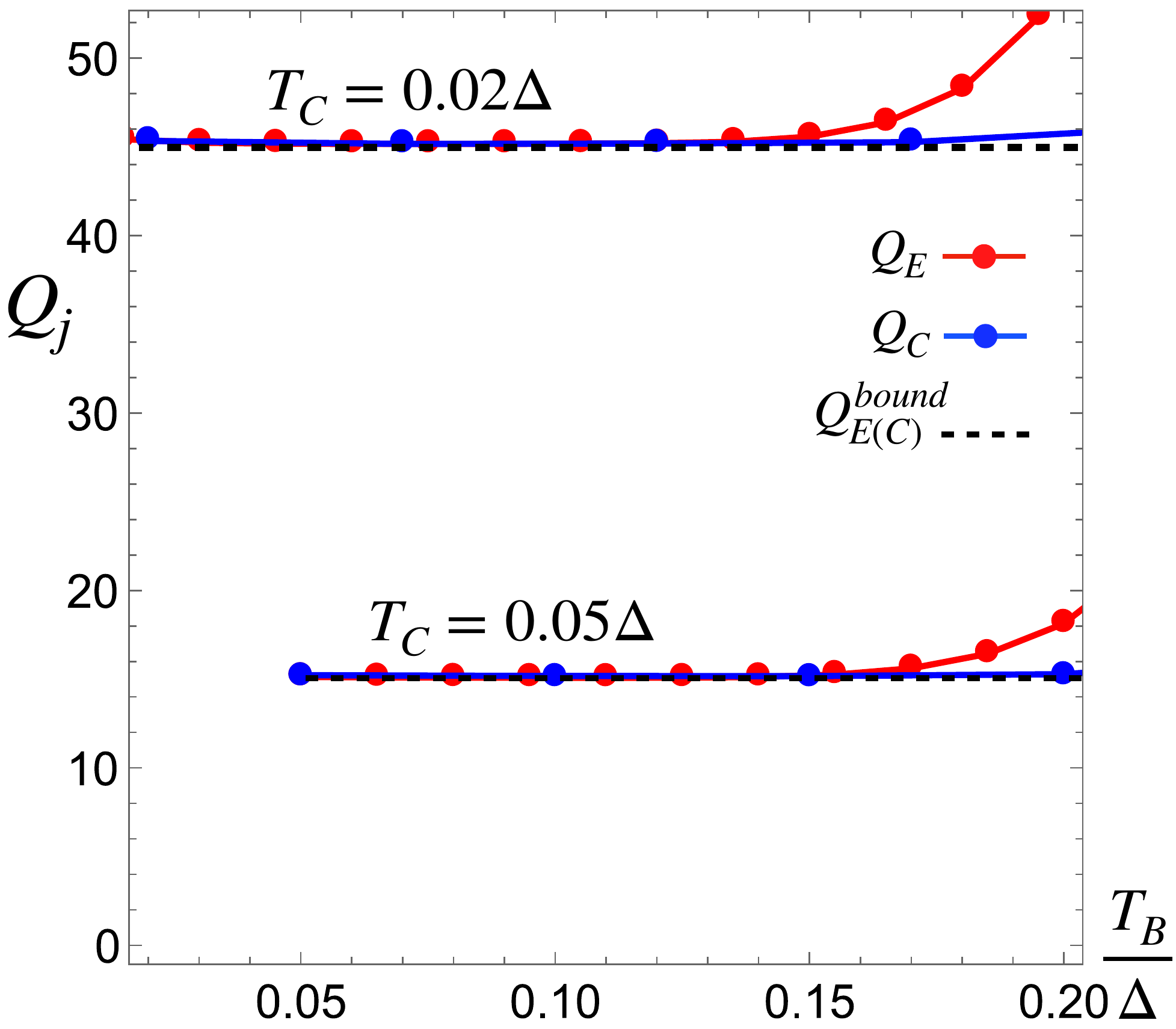}
\caption{ The plot shows $Q_J$ as a function of $T_B/\Delta$ for the unmodulated case. Here $\omega_C= \omega_0=\omega_E=0$, $\Delta=1$, and $T_E=0.2\Delta$. Here $T_B$ varies from $T_C$ to $T_E$. The lower bound of $Q_J$ is shown by the dashed black line (Cf. \eqref{turbound}). }
\label{TUR_figure} 
\end{figure}
\begin{figure}[!h]
\centering
\includegraphics[width=1\linewidth]{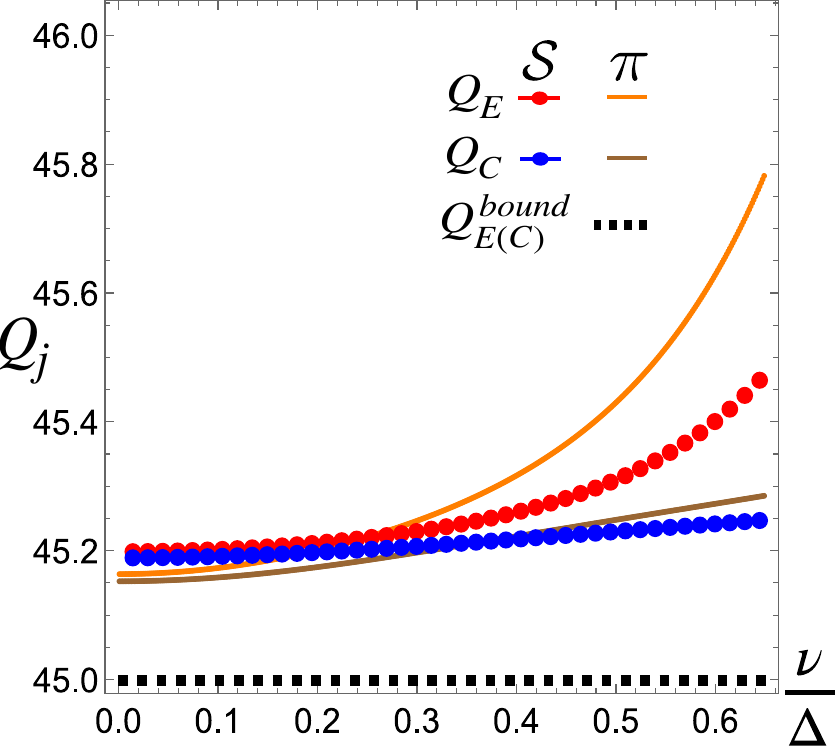}
\caption{The plot shows $Q_J$ as a function of $\nu/\Delta$ under sinusoidal and $\pi$-flip modulation for finite temperature $T_B = 0.118 \Delta$ and $\lambda = 0.8$. Here $\omega_C = \omega_0 = \omega_E = 0$, $T_E = 0.2 \Delta$, $T_C = 0.02 \Delta$ and $\Delta = 1$. Here, $Q^{bound}_{E(C)}$ represents the lower bound for both sinusoidal and $\pi$-flip modulation (Cf. \eqref{turbound}).}
\label{figTURnu}
\end{figure}
Using Eq.~\eqref{approx_sigma_eq} into Eq.~\eqref{TUR}, the TUR ratio for the emitter and the collector terminal can be simplified as
\begin{eqnarray}\label{Q-J-EC}
Q_{J_E} &\approx& F_E(\beta_C-\beta_E)= \frac{F_E}{k_B T_{eff}};\non\\ 
Q_{J_C} &\approx& F_C(\beta_C-\beta_E)=\frac{F_C}{k_B T_{eff}},
\end{eqnarray}
and are plotted in Fig.~\ref{TUR_figure} as a function of $T_{B}/\Delta$ for the unmodulated case, where $\beta_\alpha=(k_B T_\alpha)^{-1}$ and $T_{eff}=T_E T_C/(T_E - T_C)$ is some effective temperature. One can use the lower bound of Fano factor presented in Eq. \eqref{FanoBound} and the discussion below, to derive a similar lower bound in TUR, given by 
\begin{equation}
Q_{J_{E/C}}\ge Q^{bound}_{J_{E/C}} = \frac{\hbar\Delta}{k_B T_{eff}}.
\label{turbound}
\end{equation}
In order to arrive at the above bound \eqref{turbound} we have taken $\Omega_{E, C} = \Delta$ and $T_E/\Delta \gg T_C/\Delta \to 0 $ in Eq. \eqref{FanoBound}.
TUR in the presence of modulation is shown in Fig.~\ref{figTURnu}, and as shown in Figs. \ref{TUR_figure} and \ref{figTURnu}, the bound \eqref{turbound} is respected with or without modulation.
Consequently, analogous to the cases of heat engines \cite{PhysRevE.108.014137},  we observe a trade-off between precision and entropy production in the operation of quantum thermal transistors as well. We note that, in the presence of experimental constraints such as fabrication asymmetries, the degeneracy condition $\omega_E = \omega_C = 0$ may not be strictly satisfied. Nevertheless, transistor action can still be observed, as examined in detail in Ref~\cite{PhysRevLett.116.200601} and discussed in Ref.~\cite{PhysRevE.106.024110}. The effect of fabrication asymmetries or impurities, which may lead to small but non-zero deviations from the degeneracy condition ($\omega_E \neq \omega_C$), on the lower bounds of the Fano factor would be an interesting question to investigate. A detailed sensitivity analysis for the case of $\omega_E \neq \omega_C$, however, is beyond the scope of the present work.

\section{Enhancing performance of transistor through optimal control} 
\label{secV}

\subsection{Optimization protocol}
Sinusoidal and $\pi$-flip modulated transistor discussed above do not ensure optimal operation. Consequently in this section, we aim to optimize the performance of the FQT through quantum control \cite{chu02cold, chakrabarti07quantum, alessandro21introduction}, through application of the CRAB optimization protocol \cite{PhysRevA.84.022326,PhysRevLett.106.190501}. The CRAB  optimization scheme has demonstrated significant success in both theoretical \cite{mukherjee2013speeding,caneva2014complexity,PhysRevE.108.014137} as well as experimental \cite{omran2019generation,borselli2021two} studies. It is represented as a generic periodic modulation of $\omega(t)$, formulated through a truncated Fourier series: 
\begin{equation}\label{Eq:CRAB_omega}
\omega(t)=\omega_{0}+\frac{1}{2 N_c g(t)} \sum_{n=1}^{N_c}\left[a_{n} \cos \left(\frac{2 \pi n t}{\tau}\right)+b_{n} \sin \left(\frac{2 \pi n t}{\tau}\right)\right]
\end{equation}
Here  $N_c > 0$ denotes the total number of frequencies considered and $\tau$ represents modulation time period. We note that in general unbounded values of $a_n, b_n, N_c$ may be ideal for optimal control protocols \cite{PhysRevA.84.022326}. However, in view of constraints that may be encountered in experimental setups, here we choose a finite $N_c (= 10)$ and $-1 \leq a_n\leq 1$ and $-1 \leq b_n \leq 1$, which, as we show below, in general suffices to produce significant optimizations \cite{muller22one, mukherjee2013speeding}.
\begin{figure*}
\centering
\includegraphics[width=1\linewidth]{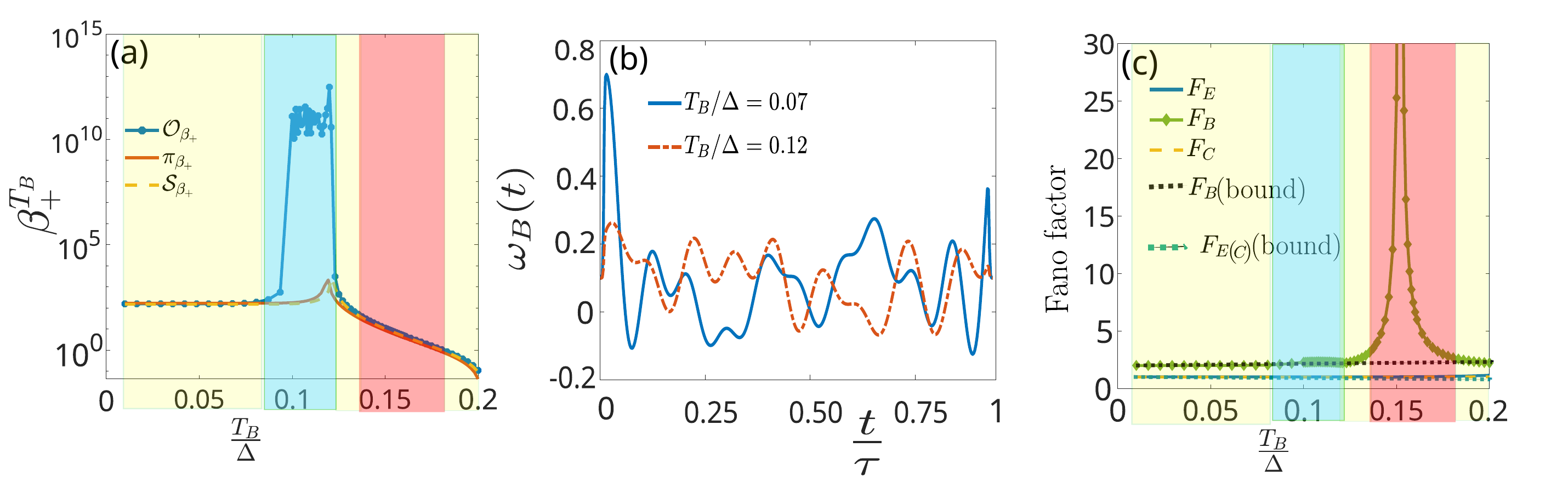}
\caption{Figure showing (a) the variation of $\beta^{T_B}_{+}$ corresponding to the optimal modulations ($\mathcal{O}_{\beta_{+}}$), sinusoidal modulation ($\mathcal{S}_{\beta_{+}}$) and the $\pi$-flip modulation ($\pi_{\beta_{+}}$) with $\frac{T_B}{\Delta}$,   (b) the optimal pulse $\omega_B(t)$ for amplification optimization using the CRAB algorithm at: (i)  $T_B/\Delta = 0.12$, which lies within the narrow regime where optimal amplification diverges, (ii) $T_B/\Delta = 0.07$, which corresponds to the regime where the optimal amplification remains nearly constant, as shown in (a), and (c) the Fano factor under amplification optimization protocol. The $\tau$ in (b) represents the total time period taken to plot the optimal modulation. The parameters used are the following: $\Delta = 1 $, $T_E= 0.2\Delta$, $T_C = 0.02\Delta$ , $\nu= 0.1$, $\omega_0 =0$.  CRAB modulation Eq.~\eqref{Eq:CRAB_omega} significantly enhances the amplification obtained for a wide range of the base terminal temperature $T_B$. We identify three distinct regimes in the parameter space relevant to the amplification–optimization protocol. 
The blue region highlights the optimal transistor regime, exhibiting high thermal amplification with minimal fluctuations (Fano factor near its lower bound)
and the red region denotes high noise and low amplification, thereby making it the least suitable regime in the parameter space for the transistor operation.  The yellow regime indicates that  operating conditions where optimal control does not have any significant effect on the output of the FQT. }
\label{fig:CRAB_Amplification}
\end{figure*}

The function $g(t)\rightarrow \infty$ for $t = 0$ and $t =\tau$, while $g(t) = 1$ for intermediate times, so that $\omega(t)=\omega_0$ at the beginning and end of a cycle. For a particular modulation frequency $\nu = \frac{2 \pi}{\tau}$, we numerically optimize the Fourier coefficients $a_n, b_n$, so as to optimize the relevant cost function, subject to certain constraints.  
In the present context, we implement the above optimization protocol to maximize the dynamical amplification $\beta^{T_B}_{+}$ of the transistor and to reduce the fluctuation of the transistor by minimizing the Fano factor $F_{\alpha}$ (see Eq.(\ref{Eq:fano})). To this end, the modulation $\omega_B(t)$ in the Hamiltonian [Eq.(\ref{EQ.1})] is taken to follow the form $\omega(t)$ as defined in Eq. (\ref{Eq:CRAB_omega}). We note that owing to the generic form of the modulation considered in Eq. \eqref{Eq:CRAB_omega}, here the current $J_{\alpha}$ and it's variance Var($J_\alpha$) are computed numerically from the generalized Lindblad matrix that includes all Floquet modes (see Eq. \eqref{Generlised_Matrix}).

\subsection{Optimizations and results}

\textit{\textbf{Amplification optimization: }}For $\omega_B(t)$ given by Eq. \eqref{Eq:CRAB_omega}, the dynamical amplification $\beta^{T_B}_{+}\approx - \beta^{T_B}_{-}$ (Eq. (\ref{EQ:5})) becomes a function of the Fourier coefficients $a_n $ and $ b_n$. We numerically optimize the coefficients $\{a_n, b_n\}$ such that the $\beta^{T_B}_{+}$ is maximized (see Fig. \ref{fig:CRAB_Amplification}a), under the application of the optimal pulse shown in Fig.~\ref{fig:CRAB_Amplification}b, for a given value of $T_B/\Delta$. 

\par 

\begin{figure}
\centering
\includegraphics[width=1\linewidth]{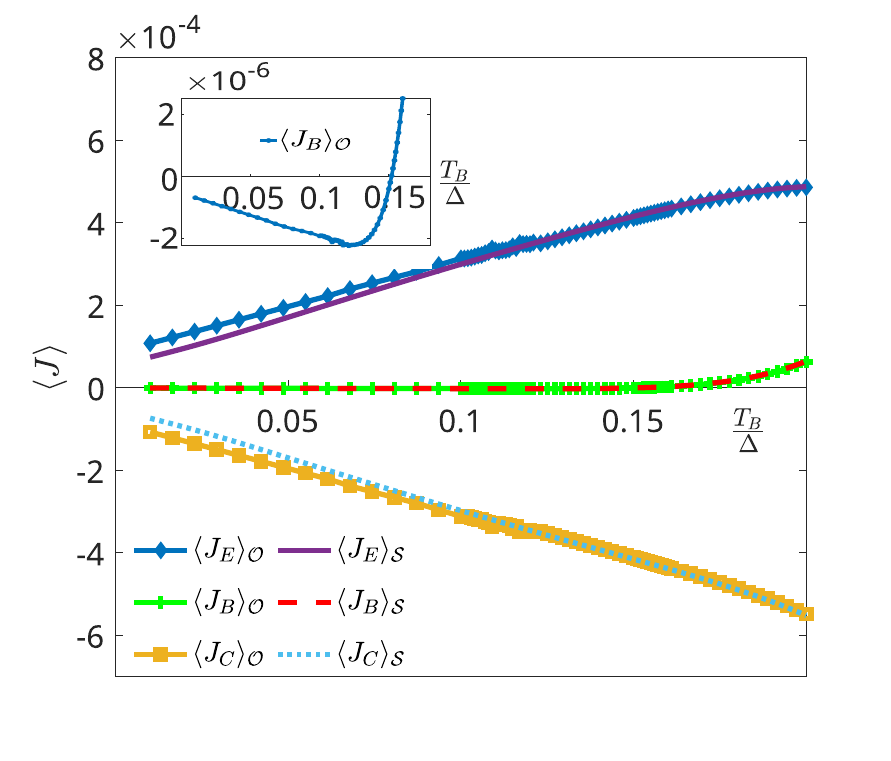}
\caption{The plot shows the variation of the currents $ \langle J_{\alpha} \rangle _\mathcal{O} $, under the optimal amplification, and the current $\langle J\rangle_{\mathcal{S}}$ for the sinusoidal modulation, with the base terminal temperature $ T_B$. The parameter used are the following: $\Delta = 1 , T_E= 0.2\Delta, T_C = 0.02\Delta , \nu=0.1, \omega_0 =0$. The inset shows the variation of $ \langle J_{B}\rangle _\mathcal{O}$ around its minimum.}
\label{fig:Current_Ampl_CRAB}
\end{figure}

As seen in Fig.~\ref{fig:CRAB_Amplification}a, the CRAB optimization protocol results in significantly higher amplifications, as compared to those obtained through sinusoidal and $\pi$-flip modulation, for a wide range of base terminal temperatures $T_B$. As expected, the obtained amplification, which is defined as the derivative of the collector/emitter current $J_{C/E}$
to the  base current $J_{B}$, shows a maximum in the range $0.1 \lesssim T_B/\Delta \lesssim 0.12$, which also corresponds to the regime of minimum $J_B $ (see Figs.~\ref{Figs_JB}a and \ref{fig:Current_Ampl_CRAB}).

Practical operation of a quantum thermal transistor demands higher amplification accompanied by low fluctuations. Therefore we present the variation of the Fano factor \( F_\alpha \) as a function of \( T_B/\Delta \) under the amplification optimization protocol, in Fig.~\ref{fig:CRAB_Amplification}c. For each value of \( T_B/\Delta \), the pulse  $\omega_B(t)$ which optimizes the corresponding $|\beta^{T_B}_{\pm}|$ is used to evaluate the Fano factor. As shown in Fig. \ref{fig:CRAB_Amplification}c,  \( F_{E/C} \) saturates close to unity under optimal amplification, whereas  \( F_B \) saturates to a value close to 2 for a broad range of \( T_B/\Delta \). Furthermore, \( F_B \) exhibits a divergence near \( T_B/\Delta \approx 0.15 \), which can be attributed to the vanishing of the corresponding base current \( \langle J_B \rangle \) at that point (see Fig.~\ref{Figs_JB}a, Fig.~\ref{fig:Current_Ampl_CRAB}).

The above numerical optimization protocol results suggest one can enhance the amplification of a FQT significantly using the optimization protocol, thereby substantially boosting its importance for application in quantum technologies. Furthermore, as shown in Figs. \ref{fig:CRAB_Amplification}, a regime around $T_B/\Delta \approx 0.1$ is ideal for operating the FQT with high amplification, yet finite fluctuations (indicated by the blue shaded region in Figs. \ref{fig:CRAB_Amplification}a and \ref{fig:CRAB_Amplification}c).   On the other hand, the regime of worst performance is indicated by the red shaded region in Figs. \ref{fig:CRAB_Amplification}a and \ref{fig:CRAB_Amplification}c, which is associated with diverging fluctuations, yet, without any substantial enhancement in the amplification.  Notably, both the above regimes are close to the parameter values which result in a minimum of $J_B$ (see Fig. \ref{fig:Current_Ampl_CRAB}), thus once more showing the non-trivial effects of a small base current.

We note that the  small oscillations around the maximum value of $\beta^{T_B}_+$ in the optimal amplification curve in Fig. \ref{fig:CRAB_Amplification}a may originate due to the existence of
multiple local maxima in the quantum control landscape \cite{alessandro21introduction, chakrabarti07quantum} of the cost function $\beta^{T_B}_{+}$, where the solution $\{a_n, b_n\}$ may get trapped, owing to the constraints considered here, viz. $|a_n|, |b_n| \leq 1$ and finite $N_c$ (see Eq. \eqref{Eq:CRAB_omega}).
The CRAB optimization scheme is expected to take us iteratively to the set of $\{a_n, b_n\}$ which results in the global maximum of $\beta^{T_B}_{+}$ with perfect accuracy only in the presence of additional resources, such as unbounded $a_n$, $b_n$, $N_c$, and randomized frequencies (see Refs. \cite{PhysRevA.84.022326, muller22one, yang25improving}). However, in practical scenarios such constraints can be expected to be an inherent part of any experimental setup. Furthermore, one can follow the above optimization protocol to maximize $\beta^{\nu}_{+}$ as well.

\textit{\textbf{Fano factor minimization: }}
Now we focus on devising an optimal pulse aimed at reducing transistor fluctuations. As observed in Fig.~\ref{F10} for the sinusoidal and unmodulated cases, and in Fig.~\ref{fig:CRAB_Amplification}c for the optimized pulse under the amplification protocol, the Fano factor associated with the base terminal is significantly higher than that of the emitter or collector terminals. Therefore, in this section we aim to minimize the Fano factor \( F_B \) (see Eq.~\eqref{Eq:fano}) associated with the base current \( \langle J_B \rangle \).
\begin{figure}[h!]
\centering
\includegraphics[width=1\linewidth]{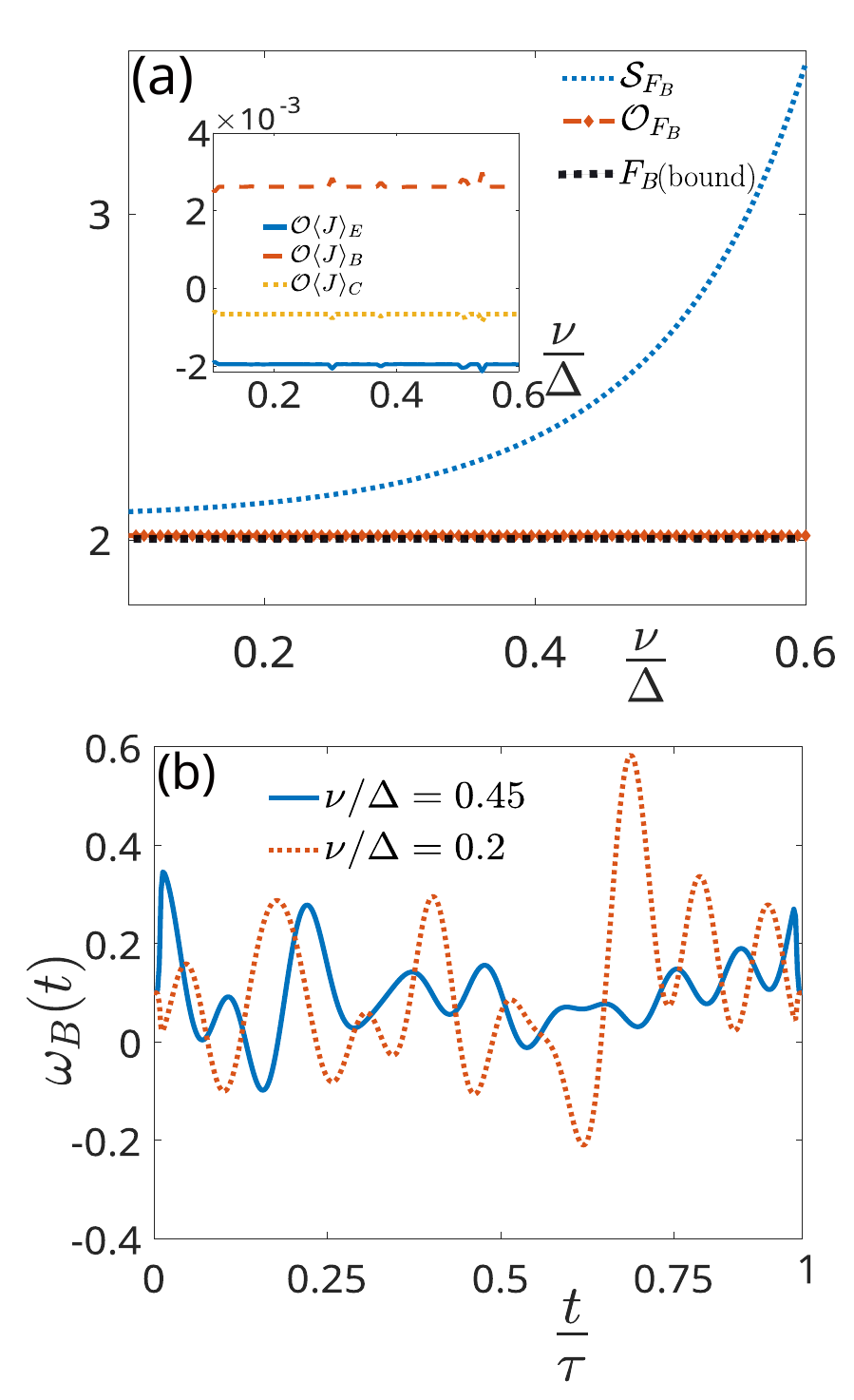}
\caption{(a) The main plot shows the variation of the Fano factor for the base terminal($F_B$) corresponding to the optimal modulation($\mathcal{O}_{F_B}$) and sinusoidal modulation($\mathcal{S}_{F_B}$) with $\nu/\Delta$.  The inset shows the variation of current $\langle J_{\alpha} \rangle_{\mathcal{O}}$ corresponding to the optimal Fano factor. A small Fano factor is associated with a large base current. (b) The plot shows the shape of the optimal pulse $\omega_B(t)$  which minimizes Fano factor, $F_B$ at: (i)  $\nu/\Delta = 0.45$ and (ii) $\nu/\Delta = 0.2$, as illustrated in Fig. \ref{fig:Optimised_Fano}(a). Here $\Delta = 1, T_B = 0.118 \Delta, T_E = 0.2 \Delta , T_C = 0.02 \Delta$.}
\label{fig:Optimised_Fano}
\end{figure}

As shown in Fig.~\ref{fig:Optimised_Fano}a,
Optimized Fano factor, $\mathcal{O}_{F_B}$ is significantly lower than the Fano factor obtained through sinusoidal modulation, $ \mathcal{S}_{F_B}$. This clearly indicates that the fluctuation in the mean base current, quantified through Fano factor, can be effectively suppressed over a broad parameter regime through suitable optimization.
However, this reduction in  $F_B$ is associated with a corresponding increase in the mean base current  $\langle J_{B} \rangle$, as seen in the inset of the Fig.~\ref{fig:Optimised_Fano}a. 
In general a thermal transistor demands $|\langle J_B \rangle| \ll |\langle J_E \rangle|, |\langle J_C \rangle|$. Consequently, in spite of the low fluctuations, the large base current obtained here can be expected to be detrimental to the operation of the device as a thermal transistor, and suggests a trade-off between the precision of the device and the amplitude of the base current. Furthermore, Fig. \ref{fig:Optimised_Fano}a emphasizes the existence of a lower bound  of $\approx 2$ for $F_B$, as discussed in Eq. \eqref{FanoBound}. The corresponding optimal pulses are shown in Fig. \ref{fig:Optimised_Fano}b.

\section{Conclusion}\label{SecVI}
We use the FCS description to study the energy exchange and fluctuations in a three-terminal FQT in the presence of periodic modulations~(see Fig.~\ref{F002} and Fig.~\ref{Figs_JB}).  
In addition, we study the Fano factor (see Fig.~\ref{F10}) and evaluate its relation with the TUR ratio (see Eq. \eqref{Q-J-EC} and Fig.~\ref{TUR_figure}), which, for weak modulation strengths, assume simple forms in the long time limit (see appendix~\ref{Appendix-A}), 
and provides valuable information about the noise characteristics of the FQT.  We exemplify our results using sinusoidal and $\pi$ pulses (see Figs.~\ref{F002} - \ref{TUR_figure}).

Finally, we use optimal control to significantly enhance the amplification obtained in the FQT (see Figs. \ref{fig:CRAB_Amplification} and \ref{fig:Current_Ampl_CRAB}). In contrast, our results suggest that optimal control may be immensely helpful for reducing Fano factor to the minimum possible value, albeit at the cost of  an enhancement in the corresponding base current, which may be disadvantageous for the operation of the device as a thermal transistor (see Fig. \ref{fig:Optimised_Fano}).   Notably,  the results obtained through the sinusoidal, pi and optimal modulations suggest existence of trade-off relations between the precision in the operation of a quantum thermal transistor, and the associated amplification and entropy production. Further, the condition  \( \omega_{E} = \omega_0=\omega_{C} = 0 \) used here, and \( \Delta \gg \omega_{B}(t)\; \forall~t \) is not the only criterion to achieve the thermal transistor effect. Even with non-degenerate 3-TLS energy levels sharing the same energy gap but smaller than the coupling energies, the transistor effect can still arise, with the crucial condition being the asymmetry between the coupling and TLS energies. In such a situation, the system reduces to at least six levels, making calculations significantly more complex~\cite{PhysRevLett.116.200601}.

We enumerate below the main results presented in this work:
\begin{itemize}
    \item We have developed a theory of full counting statistics applied to Floquet quantum thermal transistors in Sec. \ref{SecIVA}.  
    \item We have introduced  lower bounds to the Fano factors in Eqs. \eqref{f10} and \eqref{FanoBound}, which are also verified numerically in Figs. \ref{F10}, \ref{fig:CRAB_Amplification}c and \ref{fig:Optimised_Fano}a. 
    \item We have discussed the thermodynamic uncertainty relation, and introduced corresponding lower bound in Eq. \eqref{turbound}, which is verified numerically in Figs. \ref{TUR_figure} and \ref{figTURnu}. 
    \item We have studied optimal control aimed at enhancing the performance of FQTs. As shown in Fig. \ref{fig:CRAB_Amplification}, we have identified regimes of optimal performance (blue shaded regions) and of worst performance (red shaded region), through the analysis of amplification and Fano factor. 
    \item As shown in Fig. \ref{fig:Optimised_Fano}a, we have used optimal control to achieve the minimum possible Fano factor (see Eq. \eqref{FanoBound}), thereby exhibiting the immense advantage of using optimal control for significantly enhancing the accuracy of performance of quantum thermal transistors. However, this significant reduction in Fano fator is achieved at the cost of a corresponding enhancement in the base current. 
\end{itemize}

We note that controlled heat currents have already been realized experimentally in various thermal transistor setups, including on-chip superconducting qubits~\cite{PhysRevB.94.184503}, nanoscale radiative transistors~\cite{Lim2024}, vanadium dioxide based thermal transistors~\cite{10.1063/1.4952604, PhysRevLett.123.025901}, and three-terminal magnetic thermal transistors~\cite{Castelli2023}. 
Furthermore, experimental investigation of quantum fluctuations has recently been done in driven systems such as a in two-qubit NMR setup~\cite{PhysRevResearch.2.022044} and using nitrogen-vacancy (NV) centers in diamond~\cite{Hernandez2021}. For NMR-based setup, quantum state tomography was used to study precise measurements of heat exchange cumulants ~\cite{PhysRevResearch.2.022044}.
These findings open avenues for experimentally studying thermal fluctuations in Floquet quantum thermal transistors, and enhancing their performance through optimal control.  Furthermore, applications of quantum thermal transistors include efficient thermal management in nano-scale devices~\cite{Sood2018, Lim2024}, thermal switching~\cite{Castelli2023,Lim2024}, and thermal logic operations for quantum information processing~\cite{Wang2007}. Consequently, we expect our results will be of significant interest to different branches of emergent quantum technologies.

\acknowledgements
V.M. acknowledges support from Science and Engineering Research Board (SERB) through MATRICS (Project No. MTR/2021/000055) and a Seed Grant from IISER Berhampur. N.G. is thankful to IIT Kanpur for the Fellowship for Academic and Research Excellence (FARE). S.D. acknowledges support by IIT Kanpur under Grant No. DF/PDF372/2024-IITK/1344.

\onecolumngrid 

\appendix
\section{Master equation in presence of counting field }\label{Appendix-A0}
 The total density operator in the counting field evolves according to the Liouville–von Neumann equation as follows
\begin{equation}\label{Neumann}	
\frac{\partial {\rho}^{\chi}(t)}{\partial t} = \frac{i}{\hbar}\bigg(-{H}^\chi {\rho}^{\chi}(t) + {\rho}^{\chi}(t){H}^{-\chi}(t)\bigg)\end{equation}
Here,~${\rho}^\chi$ represents the counting field dependent density matrix of the composite system and \( H^{\chi}(t) = e^{-\frac{i}{\hbar}\left(\sum_{\alpha}\frac{H_{R\alpha}\chi_\alpha}{2}\right)} H(t) e^{\frac{i}{\hbar}\left(\sum_{\alpha}\frac{\chi_\alpha H_{R\alpha}}{2}\right)} \)  with  $\alpha=\{E,B,C\}$. 
In the interaction picture, the density matrix is expressed as follows
\begin{equation}
\tilde{\rho}^\chi(t)= U^\chi(t)\rho^\chi(t)U^{-\chi}(t),
\end{equation}
where $U^\chi(t)$ is the unitrary operator generated by the Hamiltonian $H(t)= H_S(t)+H_R$. The Hamiltonian in the interaction picture can be written as
\begin{eqnarray}
\tilde{H}_I^\chi(t)=U^\chi(t)H_I^\chi U^{-\chi}(t)
=\tilde{S}(t)\otimes(\tilde{\mathcal{R}}_C^{\chi_C}(t)+\tilde{\mathcal{R}}_B^{\chi_B}(t)  
+\tilde{ \mathcal{R}}_E^{\chi_E}(t)),
\end{eqnarray}
such that the equation of motion~Eq.~(\ref{Neumann})  in the interaction picture becomes
\begin{equation}\label{}	
\frac{\partial {\tilde{\rho}}^{\chi}(t)}{\partial t} = \frac{i}{\hbar}\bigg(-{\tilde{H}_I}^\chi {\tilde{\rho}}^{\chi}(t) + {\tilde{\rho}}^{\chi}(t){\tilde{H}_I}^{-\chi}(t)\bigg)
\end{equation}

Finally, under the assumption of weak coupling and  Born-Markov approximation, one arrives at the following master equation 
\onecolumngrid
\begin{equation}\label{eq6}\begin{aligned}\frac{\partial}{\partial t} \tilde{\rho}^\chi_S(t) = -\frac{1}{\hbar^2} \int_{0}^{\infty} d\tau \operatorname{Tr}_\alpha \left( \tilde{H}_I^{\chi_\alpha}(t) \tilde{H}_I^{\chi_\alpha}(t - \tau) \tilde{\rho}^{\chi}_S(t) \rho_\alpha - \tilde{H}_I^{\chi_\alpha}(t) \tilde{\rho}^\chi_S(t) \rho_\alpha \tilde{H}_I^{-{\chi_\alpha}}(t - \tau) \right.&\\ 
- \tilde{H}_I^{\chi_\alpha}(t - \tau) \tilde{\rho}^\chi_S(t) \rho_\alpha \tilde{H}_I^{-{\chi_\alpha}}(t) + \left. \tilde{\rho}^\chi_S(t) \rho_\alpha \tilde{H}_I^{-{\chi_\alpha}}(t - \tau) \tilde{H}_I^{-{\chi_\alpha}}(t) \right),\end{aligned}\end{equation}
where we have assumed \( \text{Tr}_\alpha [\tilde{H}_I^\chi ( t) \rho_\alpha] = 0 \). Now, the first term on the right-hand side of Eq.~\eqref{eq6}  can be written as
\[
\begin{aligned}
&\sum_\alpha\int_{0}^{\infty} d\tau \, \tilde{S}(t) \tilde{S}(t - \tau) \tilde{\rho}_S^\chi ( t) \Phi_\alpha (\tau), \quad \alpha=\{E,B,C\}
\end{aligned}
\]
Here, we define \( \text{Tr}_{\tilde{R}_\alpha} [\tilde{\mathcal{R}}_\alpha (\chi, t_1) \tilde{ \mathcal{R}}_\alpha (\eta, t_2) \rho_\alpha] = \text{Tr}_{\tilde{\mathcal{R}}_\alpha} [\tilde{\mathcal{R}}_\alpha (\chi - \eta, t_1 - t_2) \tilde{R}_\alpha \rho_\alpha] \equiv \Phi_\alpha (\chi - \eta, t_1 - t_2) \), with \( \Phi_\alpha (0, t) = \Phi_\alpha (t) \). Similarly, by solving the other terms, we finally obtain the following equation~
\begin{eqnarray}
\label{eq11}
\frac{\partial}{\partial t} {\tilde{\rho}}^\chi_{S} (t) &=& -\frac{1}{\hbar^2} \sum_{\{\alpha= E,B,C\}}\int_{0}^{\infty} d\tau \, [\tilde{S}(t) \tilde{S}(t - \tau) \tilde{\rho}_S^\chi (t) \Phi_\alpha (\tau)\nonumber 
- \tilde{S}(t) \tilde{\rho}_S^\chi (t) \tilde{S}(t - \tau) \Phi_\alpha (-2\chi, -\tau)\nonumber \\
&-& \tilde{S}(t - \tau) \tilde{\rho}_S^\chi (t) \tilde{S}(t) \Phi_\alpha (-2\chi, \tau) 
+ \tilde{\rho}_S^\chi (t) \tilde{S}(t - \tau) \tilde{S}(t) \Phi_\alpha (-\tau)],
\end{eqnarray}
where the correlation function $\Phi_\alpha (\tau)$ defined as  
\begin{equation}\Phi_\alpha( \tau) = \int_{-\infty}^{\infty} \frac{d\nu}{2\pi} e^{-i\nu(\chi+\tau)} G_\alpha(\nu),
\end{equation}
is related to the coupling between the bosonic reservoir and terminal \( \alpha \). Here, \( G_\alpha(\Omega_{ij}^\alpha) = G_0(\Omega_{ij}^\alpha)[1 + \bar{n}_\alpha] \).  \( G_0(\Omega_{ij}^\alpha) \) denotes the rate of spontaneous emission, while \( \bar{n}_\alpha \equiv \bar{n}_\alpha(\Omega_{ij}^\alpha) = (e^{{\hbar\Omega_{ij}^\alpha}/{k_B T_\alpha}} - 1)^{-1} \) represents the thermal occupancy of the bath mode at the frequency \( \Omega_{ij}^\alpha \) and temperature \( T_\alpha \). In the Floquet basis, $\tilde{S}(t) = \sigma_x(t)$ can be expressed in terms of $\sigma_+$ and $\sigma_-$ using the interaction picture \cite{PhysRevE.106.024110}:
$\sigma_x^B(t) = \sum_{q \in \mathbb{Z},\{\Omega_{ij}^B \}} (\tilde{\eta}(q)e^{-i(\Omega_{ij}^B+q\nu)t}\sigma_ -+ \tilde{\eta}^*(q)e^{i(\Omega_{ij}^B+q\nu)t}\sigma_+)$
~and $\tilde{\eta}(q) = \frac{1}{\tau} \int_{0}^{\tau}  e^{-i\int_{0}^{\tau} ds[\omega_B(s) - \omega_0]} e^{i q \nu t}  \, dt~$. For the emitter and collector TLS system $\sigma^\alpha_{\pm}(t)$ can be written as~\cite{PhysRevE.106.024110}:
$\sigma_\pm^\alpha(t) =  \sum_{\{\Omega_{ij}^\alpha\}} e^{\pm i \Omega_{ij}^\alpha t} \sigma_\pm^\alpha, \quad \alpha = \{E, C\} $.  After considering the Floquet basis as the eigenbasis of $\sigma_z$, we  can derive the compact expression of the Lindbadians as follows
\begin{eqnarray}\label{eq12}
\begin{split}
\mathcal{L}^{\chi_\alpha}_{\alpha}  \left[\tilde{\rho}^\chi_{S}(t)\right] =&\sum_{\{\Omega_{ij}^\alpha\}}\left[ G_\alpha(\Omega_{ij}^\alpha) \left( e^{-i \Omega_{ij}^\alpha \chi_\alpha }\sigma_- \tilde{\rho}^\chi_S(t)\sigma_+ - \frac{1}{2}\{\sigma_+ \sigma_-, \tilde{\rho}^\chi_S(t)\} \right)
\right.\\&\left.+  G_\alpha(-\Omega_{ij}^\alpha)\left(e^{i \Omega_{ij}^\alpha \chi_\alpha}\sigma_+ \tilde{\rho}^\chi_S(t)\sigma_-- \frac{1}{2}\{\sigma_- \sigma_+, \tilde{\rho}^\chi_S(t)\}\right)\right],\quad \alpha = \{E,C\}  \\
\mathcal{L}^{\chi_B}_{B}  \left[\tilde{\rho}^\chi_{S}(t)\right] =&\sum_{q \in \mathbb{Z},\{\Omega_{ij}^B \}}P_q\left[ G_B(\Omega_{ij}^B+q\nu) \left( e^{-i (\Omega_{ij}^B+q\nu) \chi_B }\sigma_- \tilde{\rho}^\chi_S(t)\sigma_+ - \frac{1}{2}\{\sigma_+ \sigma_-, \tilde{\rho}^\chi_S(t)\} \right) 
\right.\\&\left.+  G_B(-\Omega_{ij}^B-q\nu)\left(e^{i (\Omega_{ij}^B+q\nu) \chi_B}\sigma_+ \tilde{\rho}^\chi_S(t)\sigma_-- \frac{1}{2}\{\sigma_- \sigma_+, \tilde{\rho}^\chi_S(t)\}\right)\right],
\end{split}
\end{eqnarray}
In the above equation, since $\nu \ll \Omega_{ij}^B=2\Delta$, we approximate $e^{\pm i (\Omega_{ij}^B+q\nu)\chi_{_B}}$ by $e^{\pm i \Omega_{ij}^B\chi_{_B}}$ for the expression of $\mathcal{L}_B^{\chi_{_B}}$ in the main text. We emphasize that the results reported in the main text are valid for small values of $\nu$ only. On the other hand, the approximation $\nu \ll 2\Delta$ breaks down for $\nu/\Delta$ close to unity. In that case one would need to consider the exact form of the Liouvillian [Eq.~\eqref{eq12}] for evaluation of the statistics of FQTs.

\section{Constructing Liouvillian Matrix from Rate Equations }\label{Appendix-2}
The rate equations for the simplified 4-level system involving counting fields are given by:
\begin{equation}
\begin{aligned}
\frac{d \tilde{\rho}_I^\chi(t)}{dt}&= \Gamma^{\chi_E}_{_{IV\rightarrow I}}+ \Gamma^{\chi_B}_{_{III\rightarrow I}}+\Gamma^{\chi_C}_{II\rightarrow I},\\\frac{d \tilde{\rho}_{II}^\chi(t)}{dt} &= \Gamma^{\chi_E}_{_{III\rightarrow II}}+ \Gamma^{\chi_B}_{_{IV\rightarrow II}}+\Gamma^{\chi_C}_{I\rightarrow II},\\\frac{d \tilde{\rho}_{III}^\chi(t)}{dt} &= \Gamma^{\chi_E}_{_{II\rightarrow III}}+ \Gamma^{\chi_B}_{_{I\rightarrow III}}+\Gamma^{\chi_C}_{_{IV\rightarrow III}},\\\frac{d \tilde{\rho}_{IV}^\chi(t)}{dt} &= \Gamma^{\chi_E}_{_{I\rightarrow IV}}+ \Gamma^{\chi_B}_{_{II\rightarrow IV}}+\Gamma^{\chi_C}_{_{III\rightarrow IV}}.
\end{aligned}
\end{equation}
Using the above rate equations for $\tilde{\rho}_{ij}^\chi(t)$, we can construct the effective Liouvillian matrix $\mathcal{L^\chi}$ as follows: 
\begin{equation}\label{Generlised_Matrix}
\begin{pmatrix}
\mathcal{L}_{11} & \mathcal{L}_{12} & \mathcal{L}_{13} & \mathcal{L}_{14}\\
\mathcal{L}_{21} & \mathcal{L}_{22} & \mathcal{L}_{23} & \mathcal{L}_{24}\\
\mathcal{L}_{31} & \mathcal{L}_{32} & \mathcal{L}_{33} & \mathcal{L}_{34}\\
\mathcal{L}_{41} & \mathcal{L}_{42} & \mathcal{L}_{43} & \mathcal{L}_{44}
\end{pmatrix}.
\end{equation}
where,
\begin{equation}
\begin{aligned}
\mathcal{L}_{11} &=
-\Big[(\bar{n}_E+\bar{n}_C+2)\Delta
+\sum_qP_q(2\Delta+q\nu)
\big(\bar{n}_B(2\Delta+q\nu)+1\big)\Big],~~~
\mathcal{L}_{12} = \bar{n}_C\Delta e^{i\Delta\chi_C},\\[2mm]
\mathcal{L}_{13} &=
\sum_qP_q(2\Delta+q\nu)
\bar{n}_B(2\Delta+q\nu)e^{2i\Delta\chi_B},~~~\mathcal{L}_{14} = \bar{n}_E\Delta e^{i\Delta\chi_E},~~~\mathcal{L}_{21} = (\bar{n}_C+1)\Delta e^{-i\Delta\chi_C},\\[2mm]
\mathcal{L}_{22} &=
-\Big[(\bar{n}_E+\bar{n}_C+1)\Delta
+\frac{k_BT_B}{\hbar}P_0
+\sum_{q'}P_{q'}(q'\nu)\bar{n}_B(q'\nu)\Big],~~~\mathcal{L}_{23} = \bar{n}_E\Delta e^{i\Delta\chi_E},\\[2mm]
\mathcal{L}_{24} &=
\frac{k_BT_B}{\hbar}P_0
+\sum_{q'}P_{q'}(q'\nu)
\big(\bar{n}_B(q'\nu)+1\big),~~~\mathcal{L}_{31} =
\sum_qP_q(2\Delta+q\nu)
\big(\bar{n}_B(2\Delta+q\nu)+1\big)
e^{-2i\Delta\chi_B},\\[2mm]
\mathcal{L}_{32} &= (\bar{n}_E+1)\Delta e^{-i\Delta\chi_E},~~~
\mathcal{L}_{33} =
-\Big[(\bar{n}_E+\bar{n}_C)\Delta
+\sum_qP_q(2\Delta+q\nu)
\bar{n}_B(2\Delta+q\nu)\Big]\nonumber,\\[2mm]
\mathcal{L}_{34} &= (\bar{n}_C+1)\Delta e^{-i\Delta\chi_C},~~~\mathcal{L}_{41}= (\bar{n}_E+1)\Delta e^{-i\Delta\chi_E},~~~
\mathcal{L}_{42} =
\frac{k_BT_B}{\hbar}P_0
+\sum_{q'}P_{q'}(q'\nu)
\big(\bar{n}_B(q'\nu)+1\big),\\[2mm]
\mathcal{L}_{43} &= \bar{n}_C\Delta e^{i\Delta\chi_C},~~~
\mathcal{L}_{44} =
-\Big[(\bar{n}_E+\bar{n}_C+1)\Delta
+\frac{k_BT_B}{\hbar}P_0
+\sum_{q'}P_{q'}(q'\nu)
\big(\bar{n}_B(q'\nu)+1\big)\Big].
\end{aligned}
\end{equation}
In the above matrix, $q = 0,  \pm 1, \pm 2, \pm 3, \hdots$, while $q'= \pm 1, \pm 2, \pm 3, \hdots$, and $\bar{n}_{E/C}\equiv \bar{n}_{E/C}(\Delta)=\big[e^{\hbar\Delta/k_B T_{E/C}} -1\big]^{-1}$.
We are considering $\{T_E, T_B, T_C\} \ll \hbar \Delta/k_B$, which allows us to simplify the expression for the decay rates
\begin{eqnarray}
\Gamma^{\chi_E}_{_{I\rightarrow IV}}&=& \Delta\bigg(e^{-i \Delta\chi_E}\tilde{\rho}_I^\chi -e^{-{\hbar\Delta}/{k_BT_E}}\tilde{\rho}_{IV}^\chi\bigg),\non\\\Gamma^{\chi_E}_{_{II\rightarrow III}}&=& \Delta\bigg(e^{-i \Delta\chi_E}\tilde{\rho}_{II}^\chi -e^{-{\hbar\Delta}/{k_BT_E}}\tilde{\rho}_{III}^\chi\bigg), \non\\\Gamma^{\chi_C}_{_{IV\rightarrow III}}&=& \Delta\bigg(e^{-i \Delta\chi_C}\tilde{\rho}_{IV}^\chi -e^{-{\hbar\Delta}/{k_BT_C}}\tilde{\rho}_{III}^\chi\bigg),\non\\\Gamma^{\chi_C}_{_{I\rightarrow II}}&=& \Delta\bigg(e^{-i \Delta\chi_C}\tilde{\rho}_{I}^\chi -e^{-{\hbar\Delta}/{k_BT_C}}\tilde{\rho}_{II}^\chi\bigg).
\end{eqnarray}
To ensure that the system dynamics follow Born-Markov approximation, the relaxation time ($\Gamma^{-1}$) of the system  $\gg$ the characteristic timescale ($\Delta^{-1}$). Here, we take for $i>j$, in the presence of the counting field, the ratio of decay rate follows ${\Gamma^{\chi_\alpha}_{i,j}}/{\Gamma^{\chi_\alpha}_{ji}} =e^{-i \Omega_{ij}^\alpha\chi_\alpha}$, which reduces to ${\Gamma^{\chi_\alpha}_{ij}}/{\Gamma^{\chi_\alpha}_{ji}} = 1$ as $\chi_\alpha=0$.
\begin{eqnarray}
\Gamma^{\chi_B}_{_{I\rightarrow III}}&=& \sum_qP_q(2\Delta+ q \nu)e^{- 2i\Delta\chi_B}\{\tilde{\rho}_I^\chi -e^{-{\hbar(2\Delta + q\nu)}/{k_BT_B}}\tilde{\rho}_{III}^\chi\},\non\\\Gamma^{\chi_B}_{_{II\rightarrow IV}}&=&  P_0 \frac{k_B T_B}{\hbar}(\tilde{\rho}_{II}^\chi-\tilde{\rho}_{IV}^\chi)+ P_1\nu\{n_B(\nu)-n_B(-\nu)\}(\tilde{\rho}_{II}^\chi-\tilde{\rho}_{IV}^\chi).
\end{eqnarray}

\section{Characteristic Equation,  Heat Current and Amplification}\label{Appendix-A}

\begin{figure*}\centering
\includegraphics[width=1\linewidth]{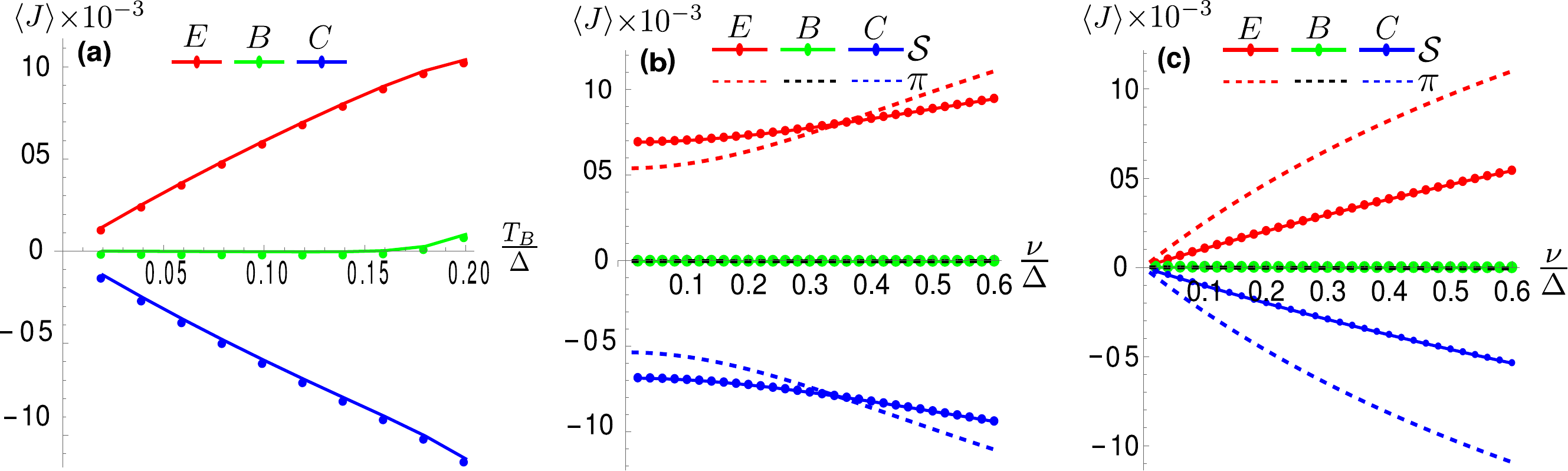}
\caption{(a) Average heat currents plot against $ T_B/\Delta$ for the unmodulated case. Parameter are $\omega_C=\omega_0=\omega_E=0$, $\omega_{EC}=0$, $\omega_{EB}=\omega_{BC}=\Delta$, $\Delta=1$, $T_E=0.2\Delta$, and $T_C=0.02 \Delta$. Fig~\ref{F01}~(b) represents the plot of average heat currents versus $\nu/\Delta$ plot under the modulated scenario. 
All parameters are the same as the Fig~\ref{F01}(a) except \(T_B = 0.118\Delta\) and \(\lambda = 0.8\). Fig~\ref{F01}(c) shows the plot heat currents against $\nu/\Delta$ for modulation case when $T_B \rightarrow0$. The parameters are the same as in Fig~\ref{F01}(b).}
\label{F01}
\end{figure*}

Starting from Eq.~\eqref{Matrix} and computing the determinant [Cf.~Eq.~\eqref{P-chi-Z}] in terms of the following quantities
\[M=e^{-\frac{\hbar\Delta }{k_BT_E}}, \quad g=e^{-\frac{\hbar\Delta }{k_BT_C}}, \quad h=e^{-\frac{\hbar\Delta }{k_BT_B}}.\] and taking the limit  $g \rightarrow 0$, we obtain the  characteristic polynomial in the following form
\begin{equation}
\begin{aligned}
P^{\chi}(z)=& (-\mathcal{F}-\Delta  (M+1)-z) \Big((\mathcal{Q}-z) (\Delta  (M (\mathcal{B}+z)+z)+z (\mathcal{B}+z)) +2 \Delta  h^2 R(\nu ) \Big((\mathcal{B}+\Delta +z) (\mathcal{Q}+2 \Delta  R(0)-z)\\
& +\Delta ^2 e^{-i \Delta  (\chi_E-2 \chi_B+\chi_C)}\Big)\Big) +\mathcal{B} \Big(-2 \Delta  h^2 R(\nu ) \Big(\mathcal{B} (-\mathcal{Q}-2 \Delta  R(0)+z)+\Delta ^2 \Big(e^{2 i \Delta  (\chi_B-\chi_C)} +M e^{i \Delta  (\chi_E-\chi_C)}\Big)\Big)\\
&+\mathcal{Q} \Big(\mathcal{B} (\Delta  M+z)+\Delta ^2 M e^{i \Delta  (\chi_E-\chi_C)}\Big) -\mathcal{B} z (\Delta  M+z)+\Delta ^2 M \Big(-2 \Delta  M R(0) e^{2 i \Delta  (\chi_E-\chi_B)} \\
&-e^{i \Delta  (\chi_E-\chi_C)} (\Delta  M+2 z)\Big)\Big) +\Delta ^2 \Big(-M \Big(2 \Delta  h^2 R(\nu ) (\mathcal{B}+\Delta +z)+\Delta  (M (\mathcal{B}+z)+z)+z (\mathcal{B}+z)\Big) \\
& +2 \Delta  h^2 R(\nu ) e^{-i \Delta  (2 \chi_E-2 \chi_B+\chi_C)} \Big(-\mathcal{B} e^{i \Delta  \chi_C}-e^{i \Delta  \chi_E} (\mathcal{B}+\Delta +z)\Big)\Big)~~.
\end{aligned}
\end{equation}
To calculate various cumulants in the long-time limit, particularly the average current and its variance, we use the following characteristic equation of the matrix $\mathcal{L}^{\chi}$ 
\begin{equation} \label{P-chi-Z}
P^{\chi}(z) = \text{det}[\mathcal{L}^{\chi}- z \mathcal{I}] \quad.  
\end{equation}
Here  $\mathcal{I}$ represents the identity matrix.  Upon expanding  the above equation in \( z \in \mathbb{R} \), we get
\begin{equation} P^{\chi}(z) = \sum_{n=0}^{N} A_n z^n= \sum_{n=0}^{N} \sum_{k=0}^{\infty} A_n^{(k)} \frac{ (i \chi)^k}{k!} z^n, \end{equation}
where $N$ is the dimension of the matrix \( \mathcal{L^\chi} \); \( A_n= \frac{\partial ^n}{\partial z^n}\frac{P^\chi(z)}{n!}\bigg|_{z=0} \) and \( A_n^{(k)} = \frac{\partial ^k}{\partial (i \chi)^k}A_n\bigg|_{\chi=0} \).  By definition, \( \tilde{\lambda}^{\chi} \) satisfies
\begin{equation} P^{\chi}(\tilde{\lambda}^{\chi})= 0 \end{equation}
which in turn leads to the following expression for the mean heat current
\begin{equation}\label{J-alpha}
\langle J_\alpha\rangle = \frac{d\tilde{\lambda}^\chi}{d(i\chi_\alpha)} \Big|_{\chi=0}= -\frac{A_0^{(1)}}{A_1^{(0)}}.
\end{equation}
Following the definition of the average heat current [Cf.~\eqref{J-alpha}], we find
\begin{equation}\label{dot1}
{\langle J_E\rangle}=\frac{\hbar\Delta ^3}{\mathbb{X}}\left[\mathcal{B} M (4 M R(0)+M+2 R(0)+2)-2h^2 R(\nu ) (-\mathcal{B} (M-3)+\mathcal{F}+\Delta  (M+2)\right]~~,
\end{equation}

\begin{equation}\label{dot2}
{\langle J_B\rangle}=\frac{4\hbar \Delta ^3}{\mathbb{X}} \left[h^2 R(\nu ) (3 \mathcal{B}+\mathcal{F}+\Delta  (M+2))-\mathcal{B} M^2 R(0)\right]~~,
\end{equation}

\begin{equation}\label{dot3}
{\langle J_C\rangle} =- \frac{\hbar\Delta ^3 }{\mathbb{X}}\left[-2 h^2 R(\nu ) (\mathcal{B} (M+3)+\mathcal{F}+\Delta  (M+2))-\mathcal{B} M (M+2 R(0)+2)\right]~~,
\end{equation}
where
\begin{eqnarray}
   \mathbb{X}&=& \Bigg[-\mathcal{B}^2 (M+2 R(0)+2)+2 h^2 R(\nu ) \left(\mathcal{B} (\mathcal{F}-\mathcal{B})+\mathcal{B} \Delta  (M+3)+3 \Delta  \mathcal{F}\right.\left.+\Delta ^2 (2 M+3)\right) +\mathcal{B} (2 \Delta \nonumber\\ &+& (M+2) (F+\Delta  M)+2 R(0) (\Delta +\mathcal{F}+2 \Delta  M))+\Delta  (M+1) (2 (R(0) (\Delta +\mathcal{F}+\Delta  M)+\mathcal{F})+\Delta  (M+2))\Bigg]~~~~~~
\end{eqnarray}
The above current expressions can be approximated as follows
\begin{equation} \label{dash1}
{\langle J_E\rangle}=\frac{\mathcal{B} \hbar\Delta ^3 M}{\mathbb{Y}} \left[4 M R(0)+M+2 R(0)+2\right]~~,
\end{equation}
\begin{eqnarray}\label{dash2}
\langle J_B\rangle = \frac{4\hbar \Delta ^3}{\mathbb{Y}} \left[h^2 R(\nu ) (3 \mathcal{B}+\mathcal{F}+\Delta  (M+2))-\mathcal{B} M^2 R(0)\right] \simeq -\frac{4 \mathcal{B} \hbar\Delta ^3 M^2 R(0)}{\mathbb{Y}}~~,
\end{eqnarray}

\begin{equation}\label{dash3}
{\langle J_C\rangle}=-\frac{\mathcal{B} \hbar\Delta ^3 M}{\mathbb{Y}} \left[M+2 R(0)+2\right]~~,
\end{equation}
where
\begin{eqnarray}
\mathbb{Y}&=&\Bigg[-\mathcal{B}^2 (M+2 R(0)+2)+\mathcal{B} (2 \Delta +(M+2) (\mathcal{F}+\Delta  M)+2 R(0) (\Delta +\mathcal{F}+2 \Delta  M))
\nonumber\\ &+&\Delta  (M+1) (2 (R(0) (\Delta +\mathcal{F}+\Delta  M)+\mathcal{F})+\Delta  (M+2))\Bigg]\simeq2( R(0)+1)\left[-\mathcal{B}^2+(\mathcal{B}+\Delta)(\mathcal{F}+\Delta)\right]~.
\end{eqnarray}
In Fig.~\ref{F01}(a), we have plotted the average $\langle J_E\rangle$, $\langle J_C\rangle$, and $\langle J_B\rangle$ against $T_B/\Delta$, with $T_E=0.2\Delta$ and $T_C=0.02\Delta$. 
Both $\langle J_E\rangle$ and $\langle J_C\rangle$ increase linearly with $T_B$ at low-temperature regime and become non-linear when $T_B \rightarrow T_E$. Consequently, a small change in $\langle J_B\rangle$ can result in significant changes in $\langle J_E\rangle$ and $\langle J_C\rangle$.

\subsubsection{Unmodulated quantum thermal transistor}
In the absence of modulation we have $\nu\rightarrow0$, and $P_q =\delta_{q,0}$. Consequently, we get $\mathcal{F}=0$,~ $\mathcal{B}={k_BT_B}/{\hbar}$,~$ R(\nu\rightarrow0)=1$, and $\mathcal{Q}=-4\Delta$ (see Eq.~\eqref{Matrix}). This yields \cite{PhysRevE.106.024110}
\begin{equation}\label{E:Q1} 
{\langle J_E\rangle}\simeq\frac{ \hbar\Delta ^2}{\mathcal{Y}_2}(\frac{k_BT_B}{\hbar\Delta}) e^{-\frac{\hbar\Delta}{k_BT_E}},
\end{equation}
\begin{equation}\label{E:Q2}
{\langle J_B\rangle}\simeq\frac{ \hbar\Delta ^2}{\mathcal{Y}_2}(\frac{k_BT_B}{\hbar\Delta})(3e^{-\frac{2\hbar\Delta}{ k_BT_B}}- e^{-\frac{2\hbar\Delta}{ k_BT_E}} ),
\end{equation}
\begin{equation}\label{E:Q3} 
{\langle J_C\rangle}\simeq-\frac{ \hbar\Delta ^2}{\mathcal{Y}_2}(\frac{k_BT_B}{\hbar\Delta}) e^{-\frac{\hbar\Delta}{k_BT_E}},
\end{equation}
where $\mathcal{Y}_2=1+(\frac{k_BT_B}{\hbar\Delta})-(\frac{k_BT_B}{\hbar\Delta})^2$.

The average heat currents result for sinusoidal and $\pi$-flip modulation are shown in Fig.~\ref{F01}(b) for $T_B=0.118\Delta$.

\subsubsection{Generic modulation with $T_B\rightarrow0$}
We now present the significant advantage of FQT in the cutoff regime (\(T_B \rightarrow 0\)), where the conventional quantum thermal transistor fails to function as a heat modulation device \cite{PhysRevE.106.024110}.  Then the expression for the heat currents can be written as follows
\begin{equation} 
{\langle J_E\rangle}\simeq\frac{ \hbar\Delta }{\mathcal{Y}_3}\nu P_1 e^{-\frac{2\hbar\Delta}{k_BT_E}}\left[ ( 4R(0)+1)+2e^{\frac{\hbar\Delta}{k_BT_E}}( R(0)+1)\right]~,
\end{equation}
\begin{equation}
{\langle J_B\rangle}\simeq\frac{4\hbar \Delta}{\mathcal{Y}_3}  \left[e^{-\frac{2\hbar\Delta}{k_BT_B}} R(\nu ) (2(2\nu P_1+\Delta))- e^{-\frac{2\hbar\Delta}{k_BT_E}} R(0)\nu P_1\right]~,
\end{equation}
\begin{equation}
{\langle J_C\rangle}\simeq-\frac{ \hbar\Delta} {\mathcal{Y}_3} \nu P_1e^{-\frac{\hbar\Delta}{k_BT_E}}\left[e^{-\frac{\hbar\Delta}{k_BT_E}}+2(1+ R(0)\right], \end{equation}
where \(\mathcal{Y}_3=2( R(0)+1)\left[2\nu P_1+\Delta)\right]\).
FQT can operate even as $T_B \rightarrow0$  with non-zero average heat currents as shown in Fig.~\ref{F01}(c).    
\begin{figure*}\centering
 \includegraphics[width=1\linewidth]{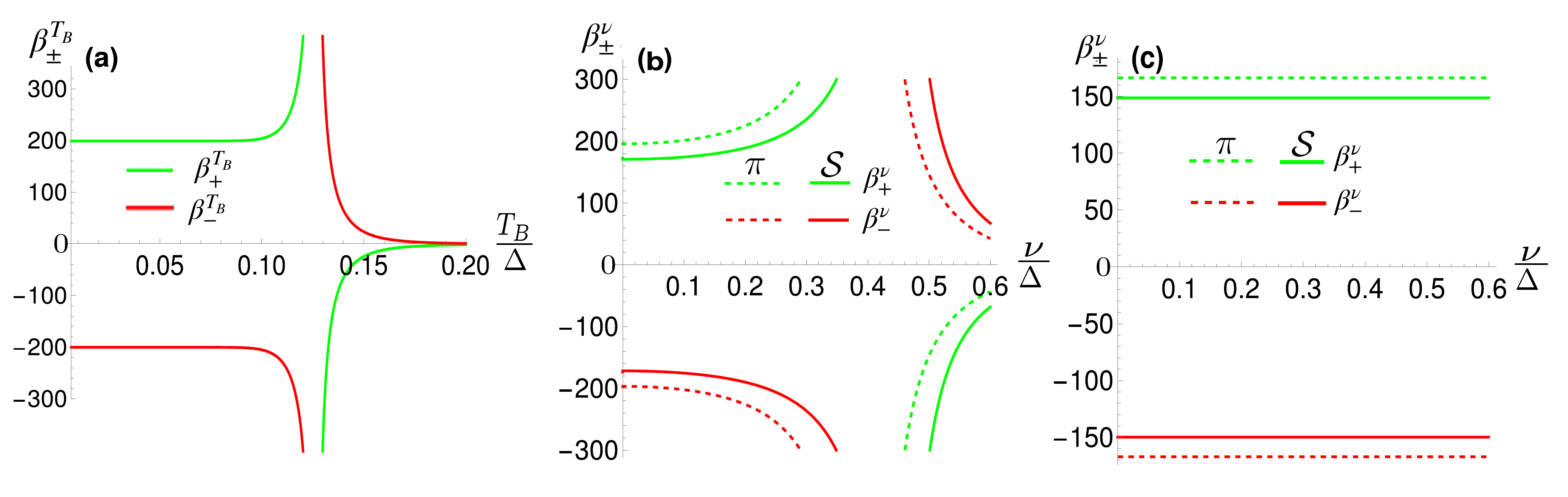}
\caption{(a)The plot shows $\beta_{\pm}^{T_B}$ against $T_B/\Delta$ for the unmodulated scenario.
Parameters include $\omega_C= \omega_0=\omega_E=0$, $\omega_{EC}=0$,$\Delta=1$, $\omega_{EB}=\omega_{BC}=\Delta$, $T_E=0.2\Delta$, and $T_C=0.02 \Delta$. Fig.~\ref{F02}(b), the dynamical amplification factors $\beta_{\pm}^{\nu}$ versus $\nu/\Delta$ for modulated scenario. The parameters are taken same as those  used in Fig.~\ref{F02}(a), except for $T_B = 0.118\Delta$ and $\lambda = 0.8$. (c)~The plot shows the dynamical amplification factor $\beta_{\pm}^{\nu}$ are plotted against $\nu/\Delta$ for modulated scenario when $T_B\rightarrow 0$.}
	\label{F02}
\end{figure*}

\subsubsection{Results of Amplification factor using counting statistics} 
 As shown in Refs.~\cite{PhysRevLett.116.200601,Guo2018,Guo2019,PhysRevE.106.024110}, the thermal current amplification factors are defined as
\begin{equation}
\beta_\pm^{T_B}
=
\frac{\partial \langle J_{C/E}\rangle/\partial T_B}
{\partial \langle J_B\rangle/\partial T_B},
\qquad
\beta_\pm^{\nu}
=
\frac{\partial \langle J_{C/E}\rangle/\partial \nu}
{\partial \langle J_B\rangle/\partial \nu}. \label{eq:amplification}
\end{equation}
From the definition of the  dynamical amplification factor~\eqref{eq:amplification}, one gets (see Eqs.~\eqref{dash1}-\eqref{dash3}):
\begin{equation}\label{beta-app} 
\beta_-^{T_B}\simeq-\frac{4 M R(0)+M+2 R(0)+2}{4 M R(0)}
~,
\end{equation}

\begin{equation}\label{BETA3}
\beta_+^{T_B}\simeq\frac{M+2 R(0)+2}{4 M R(0)}~.
\end{equation}
This amplification factor is analogous to the amplification factor of a classical transistor (Bipolar Junction transistor (BJT))\cite{PhysRevA.103.052613}.
In Fig.~\ref{F02}(a), we have shown the two amplification factors, $\beta_+^{T_B} $ and $ \beta_-^{T_B}$, plotted against temperature $ T_B$.  We observe that, at low $T_B$, $\beta $ remains significantly greater than 1. It is also noticeable that $\beta$ diverges at a certain temperature, corresponding to $T_B \approx 0.125\Delta$, where $\langle J_B\rangle$ reaches its minimum. For a tiny change in $\langle J_B\rangle$  results  substantial changes in $\langle J_E\rangle$ and$\langle J_C\rangle$, maintaining the energy conservation condition \( \sum _{\alpha = E, B, C} \langle J_\alpha\rangle = 0 \).  As  $T_B\rightarrow T_E$, the amplification factor experiences a significant decrease, dropping below unity as shown in Fig~\ref{F02}(a). Furthermore, when considering $\pi$-flip modulation the amplification factors $\beta_\pm^{\nu}$ become higher than sinusoidal modulation. This is attributed to the steeper slope of the heat currents induced by the $\pi$-flip modulation, as shown in  Fig.~\ref{F02}(b). However, as $T_B\rightarrow 0$, the absolute values of $\beta_\pm^{\nu}$ for the $\pi$-flip modulation case surpass those of sinusoidal modulation, as shown in Fig.~\ref{F02}(c). 
This observation shows the significance of $\pi$-flip modulation in improving the transistor effect under low-temperature conditions.

\providecommand{\newblock}{}


\begin{thebibliography}{10}
	\expandafter\ifx\csname url\endcsname\relax
	\def\url#1{{\tt #1}}\fi
	\expandafter\ifx\csname urlprefix\endcsname\relax\def\urlprefix{URL }\fi
	\providecommand{\eprint}[2][]{\url{#2}}
	
	\bibitem{cangemi24quantum}
	Cangemi L~M, Bhadra C and Levy A 2024 {\em Physics Reports\/} 
	\urlprefix\url{https://www.sciencedirect.com/science/article/pii/S0370157324002710}
	
	\bibitem{Bhattacharjee_2021}
	Bhattacharjee S and Dutta A 2021 {\em The European Physical Journal B\/} {\bf
		94} 239 \urlprefix\url{https://doi.org/10.1140/epjb/s10051-021-00235-3}
	
	\bibitem{PhysRevLett.116.200601}
	Joulain K, Drevillon J, Ezzahri Y and Ordonez-Miranda J 2016 {\em Phys. Rev.
		Lett.\/} {\bf 116}(20) 200601
	\urlprefix\url{https://link.aps.org/doi/10.1103/PhysRevLett.116.200601}
	
	\bibitem{bera17generalized}
	Bera M~N, Riera A, Lewenstein M and Winter A 2017 {\em Nature Communications\/}
	{\bf 8} 2180 ISSN 2041-1723
	\urlprefix\url{https://doi.org/10.1038/s41467-017-02370-x}
	
	\bibitem{saryal21universal}
	Saryal S, Gerry M, Khait I, Segal D and Agarwalla B~K 2021 {\em Phys. Rev.
		Lett.\/} {\bf 127}(19) 190603
	\urlprefix\url{https://link.aps.org/doi/10.1103/PhysRevLett.127.190603}
	
	\bibitem{PhysRevLett.2.262}
	Scovil H~E~D and Schulz-DuBois E~O 1959 {\em Phys. Rev. Lett.\/} {\bf 2}(6)
	262--263 \urlprefix\url{https://link.aps.org/doi/10.1103/PhysRevLett.2.262}
	
	\bibitem{PhysRevA.96.012125}
	Qin M, Shen H~Z, Zhao X~L and Yi X~X 2017 {\em Phys. Rev. A\/} {\bf 96}(1)
	012125 \urlprefix\url{https://link.aps.org/doi/10.1103/PhysRevA.96.012125}
	
	\bibitem{pnas.1805354115}
	Ghosh A, Gelbwaser-Klimovsky D, Niedenzu W, Lvovsky A~I, Mazets I, Scully M~O
	and Kurizki G 2018 {\em Proceedings of the National Academy of Sciences\/}
	{\bf 115} 9941--9944
	\urlprefix\url{https://www.pnas.org/doi/abs/10.1073/pnas.1805354115}
	
	\bibitem{PhysRevE.101.022127}
	Lee S, Ha M, Park J~M and Jeong H 2020 {\em Phys. Rev. E\/} {\bf 101}(2) 022127
	\urlprefix\url{https://link.aps.org/doi/10.1103/PhysRevE.101.022127}
	
	\bibitem{PhysRevB.104.075442}
	Bhandari B and Jordan A~N 2021 {\em Phys. Rev. B\/} {\bf 104}(7) 075442
	\urlprefix\url{https://link.aps.org/doi/10.1103/PhysRevB.104.075442}
	
	\bibitem{PhysRevE.106.024110}
	Gupt N, Bhattacharyya S, Das B, Datta S, Mukherjee V and Ghosh A 2022 {\em
		Phys. Rev. E\/} {\bf 106}(2) 024110
	\urlprefix\url{https://link.aps.org/doi/10.1103/PhysRevE.106.024110}
	
	\bibitem{PhysRevA.103.052613}
	Ghosh R, Ghoshal A and Sen U 2021 {\em Phys. Rev. A\/} {\bf 103}(5) 052613
	\urlprefix\url{https://link.aps.org/doi/10.1103/PhysRevA.103.052613}
	
	\bibitem{10.1063/5.0229630}
	Ekanayake U~N, Gunapala S~D and Premaratne M 2024 {\em APL Quantum\/} {\bf 1}
	036126 ISSN 2835-0103 \urlprefix\url{https://doi.org/10.1063/5.0229630}
	
	\bibitem{PhysRevB.101.245402}
	Wijesekara R~T, Gunapala S~D, Stockman M~I and Premaratne M 2020 {\em Phys.
		Rev. B\/} {\bf 101}(24) 245402
	\urlprefix\url{https://link.aps.org/doi/10.1103/PhysRevB.101.245402}
	
	\bibitem{Sothmann_2017}
	Sothmann B, Giazotto F and Hankiewicz E~M 2017 {\em New Journal of Physics\/}
	{\bf 19} 023056 \urlprefix\url{https://dx.doi.org/10.1088/1367-2630/aa60d4}
	
	\bibitem{PhysRevE.109.064146}
	Malavazi A~H~A, Ahmadi B, Mazurek P and Mandarino A 2024 {\em Phys. Rev. E\/}
	{\bf 109}(6) 064146
	\urlprefix\url{https://link.aps.org/doi/10.1103/PhysRevE.109.064146}
	
	\bibitem{PhysRevB.95.241401}
	S\'anchez R, Thierschmann H and Molenkamp L~W 2017 {\em Phys. Rev. B\/} {\bf
		95}(24) 241401
	\urlprefix\url{https://link.aps.org/doi/10.1103/PhysRevB.95.241401}
	
	\bibitem{PhysRevE.99.042121}
	Karg\ifmmode \imath \else~\i \fi{} C, Naseem M~T, Opatrn\'y T~c~v,
	M\"ustecapl\ifmmode \imath \else \i \fi{}o\ifmmode~\breve{g}\else
	\u{g}\fi{}lu O~E and Kurizki G 2019 {\em Phys. Rev. E\/} {\bf 99}(4) 042121
	\urlprefix\url{https://link.aps.org/doi/10.1103/PhysRevE.99.042121}
	
	\bibitem{PhysRevB.99.035129}
	Lu J, Wang R, Ren J, Kulkarni M and Jiang J~H 2019 {\em Phys. Rev. B\/} {\bf
		99}(3) 035129
	\urlprefix\url{https://link.aps.org/doi/10.1103/PhysRevB.99.035129}
	
	\bibitem{e24121810}
	Ghosh S, Gupt N and Ghosh A 2022 {\em Entropy\/} {\bf 24} 1810 ISSN 1099-4300
	\urlprefix\url{https://www.mdpi.com/1099-4300/24/12/1810}
	
	\bibitem{doi:10.1073/pnas.1711381114}
	Ghosh A, Latune C~L, Davidovich L and Kurizki G 2017 {\em Proceedings of the
		National Academy of Sciences\/} {\bf 114} 12156--12161
	\urlprefix\url{https://www.pnas.org/doi/abs/10.1073/pnas.1711381114}
	
	\bibitem{Ghosh2019}
	Ghosh A, Mukherjee V, Niedenzu W and Kurizki G 2019 {\em The European Physical
		Journal Special Topics\/} {\bf 227} 2043--2051 ISSN 1951-6401
	\urlprefix\url{https://doi.org/10.1140/epjst/e2019-800060-7}
	
	\bibitem{10.1116/5.0083192}
	Myers N~M, Abah O and Deffner S 2022 {\em AVS Quantum Science\/} {\bf 4} 027101
	ISSN 2639-0213 \urlprefix\url{https://doi.org/10.1116/5.0083192}
	
	\bibitem{Mandel:79}
	Mandel L 1979 {\em Opt. Lett.\/} {\bf 4} 205--207
	\urlprefix\url{https://opg.optica.org/ol/abstract.cfm?URI=ol-4-7-205}
	
	\bibitem{10.1063/1.531672}
	Levitov L~S, Lee H and Lesovik G~B 1996 {\em Journal of Mathematical Physics\/}
	{\bf 37} 4845--4866 \urlprefix\url{https://doi.org/10.1063/1.531672}
	
	\bibitem{PhysRevB.107.125113}
	Xu J, Wang S, Wu J, Yan Y, Hu J, Engelhardt G and Luo J 2023 {\em Phys. Rev.
		B\/} {\bf 107}(12) 125113
	\urlprefix\url{https://link.aps.org/doi/10.1103/PhysRevB.107.125113}
	
	\bibitem{Bouton2021}
	Bouton Q, Nettersheim J, Burgardt S, Adam D, Lutz E and Widera A 2021 {\em
		Nature Communications\/} {\bf 12} 2063
	\urlprefix\url{https://doi.org/10.1038/s41467-021-22222-z}
	
	\bibitem{Restrepo_2018}
	Restrepo S, Cerrillo J, Strasberg P and Schaller G 2018 {\em New J. Phys.\/}
	{\bf 20} 053063 \urlprefix\url{https://dx.doi.org/10.1088/1367-2630/aac583}
	
	\bibitem{engelhardt2024}
	Engelhardt G, Luo J, Bastidas V~M and Platero G 2024 {\em Phys. Rev. A\/} {\bf
		110}(6) 063707
	\urlprefix\url{https://link.aps.org/doi/10.1103/PhysRevA.110.063707}
	
	\bibitem{PhysRevB.102.195409}
	Honeychurch T~D and Kosov D~S 2020 {\em Phys. Rev. B\/} {\bf 102}(19) 195409
	\urlprefix\url{https://link.aps.org/doi/10.1103/PhysRevB.102.195409}
	
	\bibitem{PhysRevE.108.014137}
	Das A, Mahunta S, Agarwalla B~K and Mukherjee V 2023 {\em Phys. Rev. E\/} {\bf
		108}(1) 014137
	\urlprefix\url{https://link.aps.org/doi/10.1103/PhysRevE.108.014137}
	
	\bibitem{mahadeviya25current}
	Mahadeviya K, Moreira S~V, Mandal S~P, Pandit M, Prior J and Mitchison M~T 2025
	Current fluctuations in nonequilibrium open quantum systems beyond weak
	coupling: a reaction coordinate approach (\textit{Preprint}
	\eprint{2510.14926}) \urlprefix\url{https://arxiv.org/abs/2510.14926}
	
	\bibitem{1993JETPL}
	{Levitov} L~S and {Lesovik} G~B 1993 {\em Soviet Journal of Experimental and
		Theoretical Physics Letters\/} {\bf 58} 230
	
	\bibitem{PhysRevB.94.214308}
	Cerrillo J, Buser M and Brandes T 2016 {\em Phys. Rev. B\/} {\bf 94}(21) 214308
	\urlprefix\url{https://link.aps.org/doi/10.1103/PhysRevB.94.214308}
	
	\bibitem{10.1063/5.0233876}
	Zemach I, Erpenbeck A, Gull E and Cohen G 2024 {\em The Journal of Chemical
		Physics\/} {\bf 161} 164113 ISSN 0021-9606
	\urlprefix\url{https://doi.org/10.1063/5.0233876}
	
	\bibitem{PhysRevB.109.115136}
	Ding Y, Yan Y, Hu J, Engelhardt G and Luo J 2024 {\em Phys. Rev. B\/} {\bf
		109}(11) 115136
	\urlprefix\url{https://link.aps.org/doi/10.1103/PhysRevB.109.115136}
	
	\bibitem{Flindt2009}
	Flindt C, Fricke C, Hohls F, Novotný T, Netočný K, Brandes T and Haug R~J
	2009 {\em Proc. Natl. Acad. Sci. U.S.A.\/} {\bf 106} 10116--10119
	\urlprefix\url{https://www.pnas.org/doi/abs/10.1073/pnas.0901002106}
	
	\bibitem{PhysRevLett.108.183602}
	Dubost B, Koschorreck M, Napolitano M, Behbood N, Sewell R~J and Mitchell M~W
	2012 {\em Phys. Rev. Lett.\/} {\bf 108}(18) 183602
	\urlprefix\url{https://link.aps.org/doi/10.1103/PhysRevLett.108.183602}
	
	\bibitem{PhysRevE.87.012140}
	Gelbwaser-Klimovsky D, Alicki R and Kurizki G 2013 {\em Phys. Rev. E\/} {\bf
		87}(1) 012140
	\urlprefix\url{https://link.aps.org/doi/10.1103/PhysRevE.87.012140}
	
	\bibitem{chu02cold}
	Chu S 2002 {\em Nature\/} {\bf 416} 206--210 ISSN 1476-4687
	\urlprefix\url{https://doi.org/10.1038/416206a}
	
	\bibitem{chakrabarti07quantum}
	Chakrabarti R and Rabitz H 2007 {\em International Reviews in Physical
		Chemistry\/} {\bf 26} 671--735 (\textit{Preprint}
	\eprint{https://doi.org/10.1080/01442350701633300})
	\urlprefix\url{https://doi.org/10.1080/01442350701633300}
	
	\bibitem{alessandro21introduction}
	D’Alessandro D 2021 {\em Introduction to Quantum Control and Dynamics (2nd
		ed.)\/} (Chapman and Hall/CRC)
	
	\bibitem{PhysRevLett.106.190501}
	Doria P, Calarco T and Montangero S 2011 {\em Phys. Rev. Lett.\/} {\bf 106}(19)
	190501
	\urlprefix\url{https://link.aps.org/doi/10.1103/PhysRevLett.106.190501}
	
	\bibitem{PhysRevA.84.022326}
	Caneva T, Calarco T and Montangero S 2011 {\em Phys. Rev. A\/} {\bf 84}(2)
	022326 \urlprefix\url{https://link.aps.org/doi/10.1103/PhysRevA.84.022326}
	
	\bibitem{PhysRevA.87.013841}
	Shahmoon E and Kurizki G 2013 {\em Phys. Rev. A\/} {\bf 87}(1) 013841
	\urlprefix\url{https://link.aps.org/doi/10.1103/PhysRevA.87.013841}
	
	\bibitem{campbell25roadmap}
	Campbell S, D'Amico I, Ciampini M~A, Anders J, Ares N, Artini S, Auffèves A,
	Oftelie L~B, Bettmann L~P, Bonança M~V, Busch T, Campisi M, Cavalcante M~F,
	Correa L~A, Cuestas E, Dag C~B, Dago S, Deffner S, del Campo A,
	Deutschmann-Olek A, Donadi S, Doucet E, Elouard C, Ensslin K, Erker P, Fabbri
	N, Fedele F, Fiusa G, Fogarty T, Folk J~A, Guarnieri G, Hegde A~S,
	Hernández-Gómez S, Hu C~K, Iemini F, Karimi B, Kiesel N, Landi G, Lasek A,
	Lemziakov S, Lo~Monaco G, Lutz E, Lvov D, Maillet O, Mehboudi M, Mendonça
	T~M, Miller H~J~D, Mitchell A~K, Mitchison M, Mukherjee V, Paternostro M,
	Pekola J~P, Perarnau-Llobet M, Poschinger U~G, Rolandi A, Rosa D, Sánchez R,
	Santos A~C, Sarthour R~S, Sela E, Solfanelli A, Souza A~M, Splettstoesser J,
	Tan D, Tesser L, Vu T~V, Widera A, Yunger~Halpern N and Zawadzki K 2025 {\em
		Quantum Science and Technology\/}
	\urlprefix\url{http://iopscience.iop.org/article/10.1088/2058-9565/ae1e27}
	
	\bibitem{PhysRevE.75.050102}
	Talkner P, Lutz E and H\"anggi P 2007 {\em Phys. Rev. E\/} {\bf 75}(5) 050102
	\urlprefix\url{https://link.aps.org/doi/10.1103/PhysRevE.75.050102}
	
	\bibitem{Strasberg2021}
	Strasberg P 2021 {\em Quantum Stochastic Thermodynamics: Foundations and
		Selected Applications\/} (Oxford University Press)
	
	\bibitem{RevModPhys.81.1665}
	Esposito M, Harbola U and Mukamel S 2009 {\em Rev. Mod. Phys.\/} {\bf 81}(4)
	1665--1702
	\urlprefix\url{https://link.aps.org/doi/10.1103/RevModPhys.81.1665}
	
	\bibitem{PhysRevE.94.062109}
	Mukherjee V, Niedenzu W, Kofman A~G and Kurizki G 2016 {\em Phys. Rev. E\/}
	{\bf 94}(6) 062109
	\urlprefix\url{https://link.aps.org/doi/10.1103/PhysRevE.94.062109}
	
	\bibitem{Blanter2000}
	Blanter Y~M and Büttiker M 2000 {\em Physics Reports\/} {\bf 336} 1--166
	\urlprefix\url{https://www.sciencedirect.com/science/article/pii/S0370157399001234}
	
	\bibitem{Sornette2006}
	Sornette D 2006 {\em Critical Phenomena in Natural Sciences\/} 2nd ed (Springer
	International Publishing)
	
	\bibitem{Schaller2014}
	Schaller G 2014 Open quantum systems far from equilibrium {\em Lecture Notes in
		Physics\/} (Springer International Publishing)
	
	\bibitem{PhysRevResearch.5.023155}
	Prech K, Johansson P, Nyholm E, Landi G~T, Verdozzi C, Samuelsson P and Potts
	P~P 2023 {\em Phys. Rev. Res.\/} {\bf 5}(2) 023155
	\urlprefix\url{https://link.aps.org/doi/10.1103/PhysRevResearch.5.023155}
	
	\bibitem{takashi23floquet}
	Mori T 2023 {\em Annual Review of Condensed Matter Physics\/} {\bf 14} 35--56
	ISSN 1947-5462
	\urlprefix\url{https://www.annualreviews.org/content/journals/10.1146/annurev-conmatphys-040721-015537}
	
	\bibitem{alicki2012periodicallydrivenquantumopen}
	Alicki R, Gelbwaser-Klimovsky D and Kurizki G 2012 {\em arXiv\/}  1205.4552
	\urlprefix\url{https://arxiv.org/abs/1205.4552}
	
	\bibitem{mukherjee19enhanced}
	Mukherjee V, Zwick A, Ghosh A, Chen X and Kurizki G 2019 {\em Communications
		Physics\/} {\bf 2} 162 ISSN 2399-3650
	\urlprefix\url{https://doi.org/10.1038/s42005-019-0265-y}
	
	\bibitem{chen25engineering}
	Chen W, Abbasi M, Erdamar S, Muldoon J, Joglekar Y~N and Murch K~W 2025 {\em
		Phys. Rev. Lett.\/} {\bf 134}(9) 090402
	\urlprefix\url{https://link.aps.org/doi/10.1103/PhysRevLett.134.090402}

    
	\bibitem{Guo2018}
	B Q Guo, T Liu and C S Yu,
	{\em Phys. Rev. E} {\bf 98}, 022118 (2018).
		\urlprefix\url{https://doi.org/10.1103/PhysRevE.98.022118}
	
	\bibitem{Guo2019}
	B Q Guo, T Liu and C S Yu,
	{\em Phys. Rev. E} {\bf 99}, 032112 (2019).
		\urlprefix\url{https://doi.org/10.1103/PhysRevE.99.032112}
	
	\bibitem{Silaev2014}
	Silaev M, Heikkil\"{a} T~T and Virtanen P {\em Phys. Rev. E\/} {\bf 90}(2) 022103 (2014)\urlprefix\url{https://link.aps.org/doi/10.1103/PhysRevE.90.022103}
	
	

    
	\bibitem{Wagner2019}
	Wagner T, Talkner P, Bayer J~C, Rugeramigabo E~P, H\"{a}nggi P and Haug R~J
	2019 {\em Nature Physics\/} {\bf 15} 330--334
	\urlprefix\url{https://www.nature.com/articles/s41567-018-0412-5}
	
	\bibitem{PhysRevE.97.052145}
	Segal D 2018 {\em Phys. Rev. E\/} {\bf 97}(5) 052145
	\urlprefix\url{https://link.aps.org/doi/10.1103/PhysRevE.97.052145}
	
	\bibitem{PhysRevLett.133.070405}
	Guarnieri G, Eisert J and Miller H~J~D 2024 {\em Phys. Rev. Lett.\/} {\bf
		133}(7) 070405
	\urlprefix\url{https://link.aps.org/doi/10.1103/PhysRevLett.133.070405}
	
	\bibitem{koyuk22thermodynamic}
	Koyuk T and Seifert U 2022 {\em Phys. Rev. Lett.\/} {\bf 129}(21) 210603
	\urlprefix\url{https://link.aps.org/doi/10.1103/PhysRevLett.129.210603}
	
	\bibitem{Horowitz2020}
	Horowitz J~M and Gingrich T~R 2020 {\em Nature Physics\/} {\bf 16} 15--20 ISSN
	1745-2481 \urlprefix\url{https://doi.org/10.1038/s41567-019-0702-6}
	
	\bibitem{agarwalla18assessing}
	Agarwalla B~K and Segal D 2018 {\em Phys. Rev. B\/} {\bf 98}(15) 155438
	\urlprefix\url{https://link.aps.org/doi/10.1103/PhysRevB.98.155438}
	
	\bibitem{singh23thermodynamic}
	Singh V, Shaghaghi V, M\"ustecapl\ifmmode \imath \else \i
	\fi{}o\ifmmode~\breve{g}\else \u{g}\fi{}lu O~E and Rosa D 2023 {\em Phys.
		Rev. A\/} {\bf 108}(3) 032203
	\urlprefix\url{https://link.aps.org/doi/10.1103/PhysRevA.108.032203}
	
	\bibitem{zhang2025thermodynamic}
	Zhang Y and Su S 2025 Thermodynamic uncertainty relations for three-terminal
	systems with broken time-reversal symmetry (\textit{Preprint}
	\eprint{2503.13851}) \urlprefix\url{https://arxiv.org/abs/2503.13851}
	
	\bibitem{mukherjee2013speeding}
	Mukherjee V, Carlini A, Mari A, Caneva T, Montangero S, Calarco T, Fazio R and Giovannetti V 2013
	\textit{Phys. Rev. A} \textbf{88} 062326
	\href{https://doi.org/10.1103/PhysRevA.88.062326}{doi:10.1103/PhysRevA.88.062326}
	
\bibitem{caneva2014complexity}
Caneva T, Silva A, Fazio R, Lloyd S, Calarco T and Montangero S 2014
\textit{Phys. Rev. A} \textbf{89} 042322
\href{https://doi.org/10.1103/PhysRevA.89.042322}{doi:10.1103/PhysRevA.89.042322}

	\bibitem{omran2019generation}
	Omran A, Levine H, Keesling A, Semeghini G, Wang T T, Ebadi S, Bernien H, Zibrov A S, Pichler H, Choi S \textit{et al.} 2019
	\textit{Science} \textbf{365} 570--574
	\href{https://doi.org/10.1126/science.aax9743}{doi:10.1126/science.aax9743}
	
\bibitem{borselli2021two}
Borselli F, Maiw\"oger M, Zhang T, Haslinger P, Mukherjee V, Negretti A,
Montangero S, Calarco T, Mazets I, Bonneau M and Schmiedmayer J 2021
\textit{Phys. Rev. Lett.} \textbf{126} 083603
\href{https://doi.org/10.1103/PhysRevLett.126.083603}{doi:10.1103/PhysRevLett.126.083603}
	
	\bibitem{muller22one}
	Müller M~M, Said R~S, Jelezko F, Calarco T and Montangero S 2022 {\em Reports
		on Progress in Physics\/} {\bf 85} 076001
	\urlprefix\url{https://doi.org/10.1088/1361-6633/ac723c}
	
	\bibitem{yang25improving}
	Yang T, Xiong M, Xue M, Li X and Li J 2025 {\em Phys. Rev. A\/} {\bf 111}(5)
	052617 \urlprefix\url{https://link.aps.org/doi/10.1103/PhysRevA.111.052617}
	
	\bibitem{PhysRevB.94.184503}
	Karimi B and Pekola J~P 2016 {\em Phys. Rev. B\/} {\bf 94}(18) 184503
	\urlprefix\url{https://link.aps.org/doi/10.1103/PhysRevB.94.184503}
	
	\bibitem{Lim2024}
	Lim J~W, Majumder A, Mittapally R, Gutierrez A~R, Luan Y, Meyhofer E and Reddy
	P 2024 {\em Nature Communications\/} {\bf 15} 5584 ISSN 2041-1723
	\urlprefix\url{https://doi.org/10.1038/s41467-024-49936-0}
	
	\bibitem{10.1063/1.4952604}
	Ordonez-Miranda J, Ezzahri Y, Drevillon J and Joulain K 2016 {\em Journal of
		Applied Physics\/} {\bf 119} 203105 ISSN 0021-8979
	\urlprefix\url{https://doi.org/10.1063/1.4952604}
	
	\bibitem{PhysRevLett.123.025901}
	Ordonez-Miranda J, Ezzahri Y, Tiburcio-Moreno J~A, Joulain K and Drevillon J
	2019 {\em Phys. Rev. Lett.\/} {\bf 123}(2) 025901
	\urlprefix\url{https://link.aps.org/doi/10.1103/PhysRevLett.123.025901}
	
	\bibitem{Castelli2023}
	Castelli L, Zhu Q, Shimokusu T~J and Wehmeyer G 2023 {\em Nature
		Communications\/} {\bf 14} 393
	\urlprefix\url{https://doi.org/10.1038/s41467-023-36056-4}
	
	\bibitem{PhysRevResearch.2.022044}
	Pal S, Saryal S, Segal D, Mahesh T~S and Agarwalla B~K 2020 {\em Phys. Rev.
		Res.\/} {\bf 2}(2) 022044
	\urlprefix\url{https://link.aps.org/doi/10.1103/PhysRevResearch.2.022044}


\bibitem{Sood2018}
Sood A, Xiong F, Chen S, Wang H, Selli D, Zhang J, McClellan C J,
Sun J, Donadio D, Cui Y, Pop E and Goodson K E 2018
\textit{Nat. Commun.} \textbf{9} 4510
\href{https://doi.org/10.1038/s41467-018-06760-7}{doi:10.1038/s41467-018-06760-7}

\bibitem{Wang2007}
L.~Wang and B.~Li,
{\em Phys. Rev. Lett.} {\bf 99}, 177208 (2007).
\url{https://doi.org/10.1103/PhysRevLett.99.177208}

\bibitem{Hernandez2021}
S.~Hernández-Gómez and N.~Fabbri,
{\em Front. Phys.} {\bf 8}, 610868 (2021).
\url{https://doi.org/10.3389/fphy.2020.610868}
	
\end{thebibliography}
\end{document}